\documentclass[12pt]{article}
\usepackage{graphicx}
 \usepackage{longtable}
\usepackage{multirow}
\relax
\usepackage{amssymb}

\newcommand{\bo}{{\bar o}}

\def\bo{{\raise.15ex\hbox{\large$\Box$}}}               

\def\face{{\raise.2ex\hbox{$\displaystyle \bigodot$}\mskip-2.2mu \llap {$\ddot
        \smile$}}}                                      

\def\leftrightarrowfill{$\mathsurround=0pt \mathord\leftarrow \mkern-6mu
        \cleaders\hbox{$\mkern-2mu \mathord- \mkern-2mu$}\hfill
        \mkern-6mu \mathord\rightarrow$}       
\def\dvec#1{\vbox{\ialign{##\crcr
        \leftrightarrowfill\crcr\noalign{\kern-1pt\nointerlineskip}
        $\hfil\displaystyle{#1}\hfil$\crcr}}}           

\def\beq{\begin{equation}}
\def\eeq{\end{equation}}

\def\beqx{\begin{displaymath}}
\def\eeqx{\end{displaymath}}

\def\beql{\begin{eqnarray}}
\def\eeql{\end{eqnarray}}

\newcommand{\bea}{\begin{eqnarray}}
\newcommand{\eea}{\end{eqnarray}}

\newcommand{\mod}{\;{\rm mod }\;}

\def\[{\left [}
\def\]{\right ]}
\def\({\left (}
\def\){\right )}

\def\+{\oplus}

\begin{document}

\hbox{\hskip 12cm NIKHEF/2016-003  \hfil}
\hbox{\hskip 12cm Januari 2016  \hfil}

\vskip .5in

\begin{center}
{\Huge \bf Big Numbers in String Theory}

\vspace*{.4in}
A.N. Schellekens$^{a,b}$
\\
\vskip .2in

${ }^a$ {\em NIKHEF Theory Group, Kruislaan 409, \\
1098 SJ Amsterdam, The Netherlands} \\

\vskip .2in

${ }^b$ {\em Instituto de F\'\i sica Fundamental, CSIC, \\
Serrano 123, Madrid 28006, Spain} \\



\end{center}

\vspace*{0.3in}
\noindent
{\bf Abstract}\vskip .2cm
This paper contains some personal reflections on several computational contributions to what is now known as the ``String Theory Landscape".
It consists of two parts. The first part concerns the origin of big numbers, and especially the number $10^{1500}$ that appeared in our work on the covariant lattice construction  \cite{Lerche:1986cx}. This part contains some new results. I correct a huge but inconsequential error, discuss some more accurate estimates, and compare with the counting for free fermion constructions. In particular I prove that the latter only provide an exponentially small fraction of all even self-dual lattices for large lattice dimensions. The second part of the paper concerns dealing with big numbers, and contains some 
lessons learned from various vacuum scanning projects.


\vskip 1in
\noindent
\newpage

\leftline{\bf Preface (added in June 2017)}
\vskip .5truecm

This paper was written on request. On 28 september 2015 I received an invitation from Brent Nelson to contribute to a special volume of ``Advances in High Energy Physics" (Hindawi Publishing Corporation). Brent wrote: ``I'm writing to you because I have been asked to co-edit a special issue of {\em Advances in High Energy Physics} with Fernando Quevedo, Andre Lukas, Yang-Hui He and Dhaghash Mehta. The issue will be called {\em Particle and String Phenomenology: Big Data and Geometry}. The special issue will deal with all the ways we use high performance computing in addressing issues in high energy physics, and (in particular) the construction of databases of string vacua. [....]  Any original work is welcome, but so too would be any reflective piece you might compose as to the challenges of those early days (challenging in terms of computation but also in terms of the community mindset), or how you see the general `string vacuum project' unfolding." I decided to accept this invitation, and I focused in particular on the second option, a reflective piece.

A chose the title ``big numbers in string theory" but Hindawi insisted that I change the title to something more ``scientific". An unusual request, but I accepted it.
Meanwhile I learned that other authors had received unusual requests as well, such as removing some of their references. On 31 januari 2016 I received an e-mail from
Hindawi, stating  ``We regret to inform you that Special Issue has been cancelled as requested by the Guest Editors. However, we can transfer your manuscript to the regular section of the journal {\em Advances in High Energy Physics,} and it will be assigned to one of our Editorial Board Members."

I decided not to accept that offer, because I had not written this as an original research article, but as a collection of personal memories, anecdotes, a few new observations and references to some rather interesting mathematical work I had discovered during the writing.  I also decided not to submit it to any other journal. 
I believe some of this is interesting to read for a select group of people, and the arXiv is perfect for making it available to them.

\section{Introduction}


My scientific career began in 1977,  just after the start of the GUT (Grand Unified Theory) era. Undoubtedly this influenced my expectations for particle physics. 
I became convinced that
during my career I would witness major steps towards the fulfillment of  ``Einstein's dream" of deriving the laws of physics 
(in particular  the Standard Model and its parameters) from a fundamental theory.  I started working on string theory in 1985, with that long-term goal in mind. When I entered string theory it appeared to be ideally suited to realize that dream. But in 1986 I wrote a paper with Wolfgang Lerche and
Dieter L\"ust \cite{Lerche:1986cx} that radically changed my expectations.
Our  paper was  not the only one to shatter the dream of uniqueness, but such a message has more impact if 
it emerges from your own work.

During the first few months of 1987 I was deeply worried about the relevance of string theory as a fundamental theory of all interactions.  But then in the spring of 1987 I 
realized that the answer string theory gave was in fact precisely the right one to eliminate another worry that had been lingering in my mind for a few years already:
the fact that  the Standard Model seemed fine-tuned to allow interesting nuclear physics and chemistry essential for the existence of intelligent life, or at least
our kind of intelligent life. When I brought that up
in private discussions, this point of view was quickly labelled as ``anthropic", and for most people that was a synonym  for ``unscientific" or worse. Much later I learned that
Andrei Linde had come to the same insight a little bit earlier, and had even had the courage to put it in print \cite{Linde:1986fd}. Anthropic ideas were already around for more than a decade, mainly in astrophysics and cosmology, and even occasionally in particle physics, but a possible r\^ole of string theory in this story was not widely discussed until
much later, especially after Susskind's paper ``The Anthropic Landscape of String Theory" \cite{Susskind:2003kw}. 
 
I have already written extensively about this elsewhere \cite{Schellekens:2008kg,Schellekens:2013bpa}.
Here I  will not stray any further into the anthropic path (apart from some remarks in the last section)
and focus on computational issues. String theory  has driven us, whether we like it or not, in the
direction of ``Big Data". This paper is not intended as a review article, but focuses on some of my own contributions to the subject:  one of the first attempts
at estimating the size of the problem, and  work on scanning small subsets of the large number of possibilities. 
It includes some new insights gained with the benefit of hindsight.

This paper is organized as follows. In the next section I will discuss the origin of big numbers, and in particular the number $10^{1500}$ in \cite{Lerche:1986cx}. In section 
\ref{Moduli} I briefly comment on the r\^ole of moduli, which were essentially overlooked in the papers on algebraic string constructions from 1986 and 1987, but which are
now the main origin of the big numbers that characterize the landscape.
In the
second part, in section \ref{VacScan}, I discuss some attempts to handle these big numbers by means of ``vacuum scanning".    The last section contains 
a variety of thoughts concerning the r\^ole played by big numbers in the landscape.

\section{The number $10^{1500}$}

 When we wrote our paper  in 1986
we had no worries about large numbers of solutions, quite the contrary. We were not only pleased that we had found a really nice way to construct chiral string theories in four dimensions,  but
we also
believed that, for better or worse, we had understood something important, namely that in four dimensions string theory was far from unique.     To drive home that point we managed to
sneak the really big number $10^{1500}$ into our paper. This was then taken out of context and quoted enthusiastically by some, and with disgust by others. Much later,  after  
2003, we were even credited by some to have anticipated the number of flux vacua. The estimate for the latter number, $10^{500}$ 
\cite{Ashok:2003gk,Douglas:2004zg} 
was rather close to ours, with a suitable definition of ``close"\rlap.\footnote{The most recent estimate for the number of flux vacua is $10^{272000}$ \cite{Taylor:2015xtz}.}
But the only thing these numbers really have in common is that they are usually quoted out of context.

Although the proper context {\it was} provided in \cite{Lerche:1986cx}, a few additional remarks should be made, and furthermore I want to correct a huge computational error (with minor
consequences).

\subsection{The Heterotic Compactification Landscape in 1986}

But let me first sketch the scene we entered at the end of 1986.
In 1984 heterotic strings were discovered \cite{Gross:1984dd}. They gave rise to ten-dimensional chiral gauge theories. The powerful consistency conditions
of string theory seemed to determine them completely. Well, almost completely, because there were two possible gauge groups, $E_8\times E_8$ and $SO(32)$. 
Shortly thereafter it was found \cite{Candelas:1985en} that the $E_8\times E_8$ theory could be compactified on a six-dimensional Ricci-flat manifold called a Calabi-Yau manifold, which led in a rather natural way to $E_6$ GUT phenomenology. 
At that time
the belief in uniqueness was still so strong that some people were convinced that it we would soon discover why $SO(32)$ was inconsistent.  The  $E_8\times E_8$ 
theory
looked so much better  that it had to be mathematically unique. But this belief was already being challenged. 
The preprint version of \cite{Candelas:1985en} states after defining Calabi-Yau manifolds: ``Very few are known", but this phrase was removed in the published version. In June 1985
the first orbifold paper \cite{Dixon:1985jw} appeared. By now it was even more obvious that a unique outcome was not in sight, although the authors did not comment on that.

\subsection{Narain Lattice Compactification}

Then in december 1985 Narain wrote a paper entitled ``New heterotic string theories in uncompactified dimensions $<$ 10" \cite{Narain:1985jj}. Initially this was received as a bombshell, because
Narain claimed to construct entirely new string theories directly in four dimensions. 
At CERN, where I worked at the time, a leading physicist remarked ``this paper cannot be correct, because string theory is unique". The bombshell was quickly defused in 
\cite{Narain:1986am}, where it was pointed out that Narain's construction could be interpreted as a torus compactification with additional $B_{\mu\nu}$ background fields. 
Torus compactifications were already known and were not seen as a threat to the hope of uniqueness, because they only produced non-chiral theories. These could then be 
dismissed as phenomenologically irrelevant, and one could hope that one day we would find  a fundamental or dynamical reason  why chiral theories  were  selected. 
Of course, a similar problem existed with the number of space-time dimensions.  

\subsection{More Chiral Compactifications}

In Februari 1986  Strominger \cite{Strominger:1986uh} discussed  superstrings with torsion and found a large number of chiral solutions, and pointed out the urgent need for a dynamical vacuum selection principle.  Later in 1986, Kawai, Tye and Lewellen \cite{Kawai:1986va} presented their work on free fermionic constructions. They also found a large number of chiral solutions. If there was ever any hope that the requirement of chirality would lead to uniqueness, this was certainly
gone by now. 

However, we sensed an attitude of denial in the string community, and this is part of the reason why we pushed the notion that four-dimensional
string theory was not going to give rise to anything remotely like a unique answer so strongly.  We sensed this correctly. Even two decades later 
the idea that string theory was going to predict the Standard Model uniquely was common, and even nowadays the failure to meet this expectation  is sometimes used as an argument {\it against} string theory. There are no negative results in science, just bad expectations. We tried  to adjust the expectations, although only with 
 limited success.
 
 \subsection{The Covariant Lattice Construction}

Our  work can be described as a chiral version of Narain's construction. Narain compactified $p$ left-moving and $q$ right-moving chiral bosons 
on a lattice $\Gamma_{p,q}$. Modular invariance requires this lattice to be even and self-dual with respect to a metric $(+,\ldots,+,-,\ldots-)$, with
$p$ $+$'s and $q$ $-$'s. This is called a Lorentzian even self-dual lattice. The consistency condition is invariant under $SO(p,q)$, but the spectrum
is not. The spectrum depends on the norms of the left and right components of the lattice vectors, and hence is only invariant under $SO(p)$ and $SO(q)$.
This leads to a moduli space 
\beq
\label{NarainModuli}
\frac{SO(p,q)}{SO(p)\times SO(q)}
\eeq
parametrizing distinct string theories (apart from global relations). In the application to bosonic strings in $d$ dimensions, $p=q=26-d$, whereas in heterotic strings one has $p=26-d$, $q=10-d$. Generically, the moduli space has dimension $pq$, but in special points there may be additional moduli. There are always $pq$ scalars in the spectrum corresponding to the generic moduli. Their
vertex operators are, in the bosonic string, $\partial_z X^I \partial_{\bar z} X^J$, where $I$ and $J$ are the internal coordinates. In the heterotic string 
the operators are $\partial_z X^I  \Psi^J$, where $\Psi^J$ are the compactified world sheet fermions (one can get the operators in the
same form as in the bosonic string by means of picture changing).

The novelty of our approach as compared to Narain's was that we bosonized all world-sheet fermions. This allowed us to put all the momenta of the bosons on one common lattice. This even included the superghosts, and enabled us
to avoid lightcone gauge and work covariantly. Our starting point was a paper 
my collaborators had written earlier in 1986  \cite{Lerche:1986he} about modular invariance in heterotic strings using a 
covariant quantization formalism developed in \cite{Cohn:1986bn}. This led them to consider odd self-dual Lorentzian lattices\rlap,\footnote{The Lorentzian signature of that lattice should not be confused with that of the
Lorentz group nor with the signature of the Narain lattice; the lattice vectors refer only to right-moving bosons, and the negative signature belongs to the ghost components.} obtained by 
combining the roots of the Lorentz algebra with the superghost lattice. In a subsequent paper \cite{Lerche:1986ae} we found a way of replacing the odd self-dual Lorentzian lattice by an even self-dual one,
in such a way that modular invariance is maintained.

This allowed us to make use of powerful classification theorems for even self-dual lattices. In the Lorentzian case, such lattices are mathematically
unique, up to Lorentz transformations. In the Euclidean case, $p\not=0, q=0$, the  Narain moduli space (\ref{NarainModuli}) becomes trivial, but
it turns out that there exist discrete sets of lattices if $p$ is a multiple of 8. These are  referred to as ESDL's (Even Self-Dual Lattices) henceforth.
The solutions are completely known only for $p=8, 16$ and $24$, and they 
are respectively the $E_8$ root lattice; the $E_8\times E_8$ root lattice and the $D_{16}$ root lattice with a spinor conjugacy class added; and the 24
lattices of dimension 24 classified by Niemeier \cite{Niemeier}. Any attempt at guessing how this series continues will be catastrophically wrong, because
beyond 24 dimensions the number of lattices starts growing dramatically, as we will see in a moment.

\subsection{Classification of Ten-Dimensional Strings}

Having written the modular invariance condition in terms of even self-dual lattices, we were able to derive all possible 10-dimensional string
theories   with a maximal rank ({\it i.e.} 16) in ten dimensions from Niemeier's classification \cite{Niemeier}. This includes the two supersymmetric heterotic strings,
a non-supersymmetric, tachyon-free string theory with gauge group $O(16)\times O(16)$ \cite{Dixon:1986iz,AlvarezGaume:1986jb}, and five
additional tachyonic string theories, which were all already known \cite{Kawai:1986vd}. 
Lattice methods are limited to maximal rank by construction. 
In ten dimensions there is one heterotic string theory that cannot be obtained that way, with gauge group $E_8$, 
realized as an affine algebra at level 2 \cite{Kawai:1986vd,Bennett:1986et}. This additional theory can be obtained from a generalization of the 
Niemeier lattices to conformal field theory. The number of such theories has been conjectured to be 71 \cite{Schellekens:1992db}, and if this list
is indeed complete, it implies the completeness of the list of heterotic strings in ten dimensions. 

\subsection{World-Sheet Supersymmetry}

After completing \cite{Lerche:1986ae} we tackled the analogous problem in four dimensions. There was just one hurdle to be taken, namely to find a way of realizing world-sheet supersymmetry. This was non-trivial since all word sheet fermions had been bosonized. But we could adopt a solution to that problem already
proposed by Kawai et. al. \cite{Kawai:1986va} (after our paper was written we learned that Antoniadis, Bachas, Kounnas and Windey \cite{Antoniadis:1985az} had already
presented this solution in october 1985, in a paper not only missed by us but apparently also by Kawai et. al.).  This solution amounted to requiring the
presence of a definite set of norm-4 vectors  on the right-moving part of the lattice. With that ingredient added, the rest of the work was so straightforward that one of us even wondered if it was worth writing a paper about it. But there was little time to worry about that. A month earlier Kawai et. al. had published a second paper in which
they introduced ``charge lattices" derived from their free earlier fermion construction. These are odd self-dual lattices,  clearly related to the covariant lattice construction we had already
introduced and applied in  our work on ten-dimensional strings \cite{Lerche:1986ae}. We were convinced that our even self-dual lattice formulation was superior in
elegance, and that therefore a four-dimensional sequel to \cite{Lerche:1986ae} {\it was} worth writing. And there was more competition putting us under pressure to
act quickly:
while we were writing our paper we heard that Antoniadis, Bachas and Kounnas were working on free fermionic constructions of chiral four-dimensional strings.
We submitted our paper on 24 november 1986, and it was followed just before the end of the year by the ABK paper  \cite{Antoniadis:1986rn}. Shortly after that 
Narain, Sarmadi and Vafa published a paper on yet another construction they called ``asymmetric orbifolds" \cite{Narain:1986qm}.
During the year 1987 it became clear that all these different constructions are closely related.  However, the relations are 
all based on studies of classes of examples; even today
a satisfactory overall picture is still missing.

The final result of our work was a class of string theories described by even self-dual lattices $\Gamma_{22,14}$, with the additional requirement that the right-moving
part of the lattice was build out of the weight lattices of $D_5 \times (D_1)^9$ with a definite set of conjugacy classes required to be present
in order to have world-sheet supersymmetry. These classes are (here $v$ denotes the vector conjugacy class of the orthogonal lattices)
\begin{eqnarray}
\label{TripletConstraint}
(v,v,v,v,0,0,0,0,0,0)  \nonumber \\ 
 (v,0,0,0,v,v,v,0,0,0)\\
 (v,0,0,0,0,0,0,v,v,v)\nonumber
\end{eqnarray}
Modular invariance
at arbitary genus\footnote{Assuming that the usual problems associated with the superghost partition function at arbitrary genus can be overcome.}  is guaranteed by the even self-duality of the lattice.

\subsection{Lattice Classification Theorems}

We realized that the rigid structure of the right-moving part of the lattice freezes the Lorentz rotations of  Narain compactifications. Hence we
expected a discrete set of solutions, rather than the continuous Narain moduli space. But how could this class be enumerated? At this point we used a trick. Under modular transformations, the characters of a right-moving $D_n$ factor transform in the same way as a left-moving $D_{8-n}$ factor. This
allowed as to replace the right-moving $D_5 \times (D_1)^9$ lattice factor by a left-moving $D_3 \times (D_7)^9$, which gives rise to a Euclidean lattice
of total dimension 88. Hence all solutions can be read off from a list of such lattices. But we quickly realized that this was totally useless as an
approach to enumeration, because the list of such lattice is unfathomably large. 

There is an amazing formula known as ``the Siegel mass formula" (see \cite{Conway259}), which states
\beq
\label{Siegel}
\sum_{\Lambda} \frac{1}{g(\Lambda)} = \frac{B_{4k}}{8k} \prod_{j=1}^{4k-1} \frac{B_{2j}}{4j} \equiv L_{8k} \ ,
\eeq
where $B_{2j}$ are the Bernoulli numbers, the number of dimensions is $8k$, and $g(\Lambda)$ is the order of the discrete automorphism group of $\Lambda$. Since this group
has at least two elements (the identity and reflection through the origin), one derives a lower limit on the number $N_{8k}$ of lattices in $8k$ dimensions:
\beq
\label{SiegelTwo}
N_{8k} > 2L_{8k}
\eeq 
In \cite{Lerche:1987sk} we made a quick estimate of $L_{88}$, and found a number of order $10^{1500}$. The actual number of lattices would have to be larger still, but the restrictions to be imposed on them (namely the presence of a $D_3 \times (D_7)^9$ and the vectors (\ref{TripletConstraint})) will reduce the total by another huge factor.

\subsection{Finiteness, free fermions and enumerability}

Although the map to Euclidean lattices is  useless as a path towards enumeration,  it did serve another purpose, namely showing
that the number of solutions is finite. Finiteness is obvious in free fermion constructions with (anti)-periodic boundary conditions.  If the fermions are complex\footnote{Real fermions allow for additional possibilities that cannot be written in terms of lattices.}, the resulting four-dimensional string theory can be written in terms of a lattice. 
But the converse is not true:
Most lattices cannot be written in terms of free fermions. If the lattice is made out of $D_n$ roots and weights, one obviously can, but not if it is made out of $A_n$ roots and weights. For example, the $A_2$ weight lattice defines a free bosonic CFT with conformal weights $0$ and $\frac13$, while in free fermionic CFT's
one can only get multiples of $\frac18$.

However, we are considering even self-dual  lattices here, where all conformal weights are integer. 
Often in string theory two construction are related even if the
relation is not manifest. So it is certainly imaginable that all self-dual lattices can be written in terms of free fermions, even if they are built out of $A_n$ lattices or
other  factors. In that case, the fermionic construction would provide a proof of finiteness. However, this does not work. In the appendix we  show that there exist
Euclidean even self-dual lattices that cannot be written in terms of free fermions with (anti)-periodic fermions. 
Hence the Siegel mass formula provides the only presently known way to demonstrate that the number of lattice solutions is finite. Indeed, I do not
 no know an  algorithm that
would generate all even self-dual lattice of a given dimension in a finite amount of time. 

One path towards such an algorithm might be the following. 
One starts by decomposing the $8k$-dimensional lattice into $8k$ $U_{2R}$ factors. Here $U_{2R}$ is a compactified chiral boson with
$2R$ primaries. Hence $U_2$ is equivalent to $A_1$ level 1 and $U_4$ is $D_1$, equivalent to a complex pair of free fermions. Such a decomposition should always be possible due to the lattice structure. The lattice CFT has $8k$ free bosons, and the only allowed operators in such a CFT are derivatives of the bosons
multiplied with $e^{i\vec v X}$, where $v$ is a lattice vector. The fact that the unit cell of a self-dual lattice is non-zero implies that a basis must exist such that
all components $v_i$ are quantized. 

For given $R$ this yields a finite algorithm for the construction of all even self-dual lattices of  a given dimension.
However it is not clear what value of $R$ should be used for given $k$. For $k=1$ and $k=2$ the value $R=2$ suffices (indeed,
$R=1$ is already enough) but in the appendix we show that for $k > 8$ $R=2$ is not large enough (presumably $R=2$ is insufficient already for smaller values 
of $k$). 
What is missing to turn this into a finite algorithm is an upper limit of $R$ for given $k$. It is clear that this upper limit increases with $k$. The same argument used in the appendix to estimate the number of fermionic theories can be used for products of $U_{2R}$. For given $R$, the upper limit for the
number of 
such lattices is given by $f(R)^{k^2}$, where $f(R)$ is a large integer that depends on $R$, but not on $k$ (for $R=2$, the most conservative upper limit
derived in the appendix is $f(2)=4^{64}$). Hence the upper limit only grows with a power $k^2$, which for any value of $R$ is always surpassed by $L_{8k}$, for sufficiently large $k$. 

\subsection{An Upper Limit?}

In stating that the total number of lattices is finite we made an assumption, namely that the group of discrete symmetries of a finite-dimensional lattice is always finite. There is in fact a plausible candidate for the $8k$-dimensional lattice with the maximal number of discrete symmetries, namely the $D_{8k}$ root lattice. This can be made self-dual by adding a spinor conjugacy class. In eight dimensions this enhances the discrete symmetries: one obtains the root lattice of $E_8$, and hence the discrete symmetries are the Weyl group of $E_8$, which is larger than the Weyl group of $D_8$. This happens because the additional spinor roots provide additional Weyl reflections. 
 In 16 dimensions and more this enhancement does not occur, and the
order of the symmetry group of the root lattice of $D_{16}$ with a spinor conjugacy class added is that of the Weyl group of $D_{16}$. However, one may check that 
the discrete symmetry group of the $E_8 \times E_8$ lattice (this group is the product of the two Weyl groups and the interchange symmetry of the factors) is larger than the
Weyl group of $D_{16}$.

In 24 dimensions it is $D_{24}$ that provides the largest set of discrete symmetries, and it is easy to check that in $8n$ dimensions
for $n > 3$ the Weyl group $W_{8k}$ of $D_{8k}$ is larger than the symmetry group of the lattice $(E_8)^k$.  In general, symmetry groups of lattices 
always contain the Weyl group of the root system, but usually this is only a small subgroup of the full automorphism group. Indeed, the Leech lattice
has no roots at all, but its automorphism group is a double cover of the Conway group, and has order $8,315,553,613,086,720,000 \approx 8.3 \times 10^{18}$.
However, the Weyl group of $D_{24}$ has order $2^{23} (24!) \approx 5.2 \times 10^{30}$, which is far larger. It is known that beyond 24 dimensions the bulk of
the lattices has  a root lattice of rank less than $8k$, and there are huge numbers of ESDL's with no roots at all. 
With only one specimen of rootless  lattices known explicitly it is a bit hazardous to speculate, but one might
conjecture that the Weyl group of $D_{8k}$ is the largest
possible lattice symmetry group for all $k>3$. Assuming this is true, we get the inequality
\beq
\label{SiegelThree}
N_{8k} <  |W_{8k}| L_{8k}
\eeq 

\subsection{Upper and Lower Limits for $8k \leq 88$}

It is straightforward to compute the number $L_{8k}$ exactly. We have computed  these numbers for all dimensions up to $88$, and they are shown (multiplied by 2) in the second column of table \ref{BigNumberTable}. The conjectured upper limits are shown  in 
column 3 of the table, where of course for $k=1$ and $k=2$ we used $E_8$ and $E_8\times E_8$. 

The estimate ``$10^{1500}$" of \cite{Lerche:1986cx} was supposed to be an approximation of the last number in the second column of the table. Hence the exact computation
yields about $10^{930}$ instead of $10^{1500}$.
Sadly,  this was one of the worst estimates in the history of science. This is presumably due to an algebraic error rather than
a poor approximation. Indeed, there exist extremely accurate estimates of the Bernoulli numbers (see eqn. (\ref{StepSize}) in the appendix) which we presumably used.

But this is still gives only a lower limit on the number of lattices.  
The true upper limit, assuming a maximal automorphism group, is about $10^{1090}$, as shown in the
third column of the table.  
None of this makes any difference for the conclusions, 
however. The only scientific point we were making is that the number is finite and large, and that is true regardless of the estimate. 
In six and eight space-time dimensions the same arguments can be used, and the Euclidian lattice dimension is
respectively 64 and 40. We read off from the tabel that there are at most $10^{455}$ resp. $10^{111}$ such lattices.

\vskip .3truecm
\begin{table}[h!]
\begin{center}
\begin{tabular}{|c|l|l|c|c|c|} \hline
Dimension &  Lower limit & Upper limit & Actual Number   \\ \hline \hline
8  &  $2.870554085831864 \times 10^{-9}$ & 1& 1 \\
16  & $4.977181647677474 \times 10^{-18}$  & 2.4160839160839161& 2 \\
24  &  $1.587356093933540 \times 10^{-14}$ & $4.130854882089717 \times 10^{16}$ & 24 \\
32  &  $8.061846587120415 \times 10^{7}$ & $2.277750478211998 \times 10^{52} $ & ? \\
40  &  $8.786162893954708 \times 10^{51}$  & $1.970535004851803 \times 10^{111}$& ? \\
48  & $3.051507011767375 \times 10^{121}$   & $2.665648986868395 \times 10^{196}$& ? \\
56  &  $1.276238439666753 \times 10^{219}$  & $1.634633237068218 \times 10^{310}$ & ? \\
64  &  $9.544539505706936 \times 10^{346}$  & $5.585108422305436 \times 10^{454}$ & ? \\
72  &  $9.613130349683812 \times 10^{506}$  & $6.949609601107582 \times 10^{631}$& ? \\
80  &   $5.862018298127880 \times 10^{700}$ & $1.267986279010439 \times 10^{843} $ & ? \\
88  & $6.485314719426174 \times 10^{929} $   & $9.307090939221263 \times 10^{1089}$& ? \\
\hline
\end{tabular}
\caption{Minima and maxima of the number of even self-dual lattices in $8k$ dimensions.\label{BigNumberTable}}
\end{center}
\end{table}

\subsection{Dimension 32}

A bit more is known for dimension 32. The Siegel mass formula can be worked out for any given root lattice 
separately, and this was done in \cite{MR1954971}. The root lattice of an ESDL is by definition the sub-lattice spanned by the vectors of norm 2. This may be trivial, and it may
be of lower dimension than the lattice under consideration. If the rank of the root lattice is equal to the rank of the ESDL, the root lattice is called {\it complete}.
Summing over the contributions of all 13218 root lattices that can actually occur (including the trivial class with no  roots at all), one obtains a lower bound of $1.1 \times 10^9$. 
This  involves a substantial amount of computation. 
To find out which
root lattices can occur one works out all combinations of simple Lie algebra root lattices with rank 32 or less, and works out the Siegel mass formula of that class. If it is zero, there are no  lattices  with that
root lattice, and if it is non-zero, there must be at least one. 

This analysis shows also that there are at least $10^7$ ESDL's of dimension 32 without roots. Furthermore it
is known that there are exactly 132 even selfdual lattices that are indecomposable and complete \cite{MR1279061}.  This means that their root lattice,
generated by the vectors of norm 2,  has rank 32 but is not a direct sum of lattices of dimension 8 and 24 or twice 16.  These 132 ESDL's have 
119 different root lattices. If we include $(D_{16})^2$ and $E_8$ times the 23 Niemeier lattices with a complete root system, we get a total of 143 distinct complete root systems,
a small fraction of the full set of 13218. 

One could use also this approach to work out an upper bound along the lines sketched above. In the tex source of \cite{MR1954971} there is a table
giving for all 13218 root lattice the Siegel mass (the sum of reciprocals of orders of automorphism groups) times the order of the Weyl group. 
Let us call this number ({\it i.e.} Siegel mass times Weyl group order) $m_R$, where $R$ labels the root lattices. Instead of formula (\ref{SiegelThree}) we 
now get
\beq
\label{SiegelFour}
N_{8k} <  \sum_R m_R \frac{|W_{8k}|}{|W_R|}\ ,
\eeq 
where $W_R$ is the Weyl group of $R$. This sum can be worked out explicitly using the  list in \cite{MR1954971}, but it is not going to improve 
the upper bound in the table by much, if anything, because there are several individual contributions of order a few times
$10^{51}$.

\subsection{A More Accurate Estimate}

A more accurate estimate for the number of covariant lattices in four dimensions can be made along the lines
of the arguments in the appendix, at least for the subset that can be realized in terms of free fermions. 
This means
that although the lattices we need are generically of the form $D_3 \times (D_7)^9\times \Gamma_{22}$, 
we will only consider the subset where $\Gamma_{22}$ is
a product of 22 $D_1$ factors. There may be additional lattices, but since we have only 22 dimensions that can deviate from free fermions, we can try to
estimate what is missing by asking which fraction of the 24 Niemeier lattices cannot be realized in terms of free fermions. This fraction may well be zero, but at least
eight of the Niemeier lattices (including the Leech lattice) have an explicit free fermion realization, so in that case fermionic constructions are underestimating
the total by at most a factor 3. The fraction of non-fermionic lattices increases rapidly with dimension, so since 22 is close to 24, but smaller, this suggests that we
will not be missing too much  -- at least on a log scale --  by limiting ourselves to free fermions.

Now we have a lattice with $32$ orthogonal group factors, and this problem is combinatorially analogous to $(D_1)^{32}$, as was already pointed out
in a footnote in \cite{Lerche:1986cx}.
This implies that one can bring the lattice basis to the form shown just above Eqn. (\ref{TNMEK}) in the appendix. This basis divides the 32 lattice components
into three blocks, one of size $N$, a second one of size $32-2N$ and a third one that is again of size $N$.

Since  the 32 orthogonal factors are not all identical one has to sum over all
possible ways of distributing $D_3$, $D_7$ and $D_1$  over the three blocks, which enlarges the total number of distinct possibilities. For each such choice the total number of possibilities is given by the formulas given in the appendix, such as Eqn. (\ref{LatticeEstimate}). This estimate gives a good approximation to the number of
lattices a computer would find in a systematic search. The only source of inaccuracy is due to the estimate of the fraction of vectors that have a certain required length (modulo 2) and a certain inner product (modulo 1). Since we are now dealing with mixed $D_1$, $D_3$ and $D_7$ lattices, the inaccuracy is
harder to estimate than in the appendix. 
The easiest way to deal with this is take into account the
maximal observed variation in both directions as an error estimate. For  lattices of 32 components that gives an error of about $\pm 3$ in the exponent.

Now we have to impose the condition that the constraint vectors (\ref{TripletConstraint}) are on the lattice. Since the lattice is self-dual, this is true
if and only if the constraint vectors have integer inner product with all lattice vectors. To check this, we have to consider the inner product of each of these
three vectors with the lattice spinor generators, {\it i.e.} the first $N$ glue vectors. The inner product can be integer or half-integer, with each possibility
occurring roughly in half of the cases. Hence we get a reduction by a factor $2^{-3N}$. Although the $D_7$ factors are associated with definite lattice factors,
there may be several distinct choices for combining them into triplets. These choices must also be summed over. The total number of distinct covariant lattice choices 
$N_{\rm CL}$ is then
\begin{equation}
N_{\rm CL}=\sum_{N=0}^{16}  d(N) 2^{32(N-1)-N^2-3N}\ ,
\end{equation}
where $d(N)$ is the number of distributions of $D_3$ and  $(D_7)^9$ over the three lattice blocks generated by the basis choice and
the number of distinct assignments of $D_7$ to triplets. Note that for some assignments, for example when all of $D_3 (D_7)^9$ is in the first $N$ components,
no choice of constraint vectors is on the lattice. Then this configuration does not contribute to $d(N)$. The general rule is that each of the three constraint
vectors must have at least one component in the second or third block, so that their inner products with the $N$ spinor classes can be tuned to the
the correct value. This is because the first $N$ entries of the spinor conjugacy classes are already fixed.

The factors $d(N)$ can be computed exactly, and are of order a few tens to at most  few hundreds.  The exponential factor takes its maximal value for $N=14$ and $N=15$, and this value is about $10^{53}$. Taking into account the error in estimating inner products, and allowing two orders of magnitude for the estimate of 
$d(N)$,
this gives an upper limit of about $10^{55\pm 5}$. This is the maximal number of orthogonal group lattices that can be generated by a computer going systematically through all possibilities.  Of course this number is still reduced by lattice degeneracies, which potentially gives a huge suppression. This suppression is less than it is
for $(D_1)^{32}$, because not all factors are identical. Based on what we know about lattices of dimension 32 and less, it may be expected to reduce the number by 20 to 50 orders of magnitude. The total number of distinct four-dimensional covariant lattice theories with the triplet supercurrent then comes out somewhere between a mere $10^5$ to $10^{35}$,  with large uncertainties, but
in any case nowhere near $10^{1500}$.  But this number was never claimed  as an estimate anyway.

Using the method of 
\cite{MR1954971}
mentioned above one should be able to find lower bounds on the number of lattices of dimension 88 with a root lattice $D_3 \times (D_7)^9 \times X_{22}$ where $X_{22}$ is
any root lattice of dimension 22 or less. This would undoubtedly be a considerable amount of work but it might be doable, since the number of options for $X_{22}$ is 
a subset of all possible 32 dimensional root lattices, which have already been enumerated. If these results could be extended to lattices generated by vectors of norm 4 one might be able to able to get lower bounds on the number of lattices containing the supersymmetry constraint vectors (\ref{TripletConstraint}), {\it i.e.} the ones of actual physical interest. But this sounds to good to be true.

 \subsection{Other supercurrents}

There are more four-dimensional strings with a covariant lattice description than the ones described above.
The choice of supercurrents (\ref{TripletConstraint}) is not the only possible one. A second possibility is to use
nine vectors of the form $(v,0,\ldots,\sqrt{3},\ldots,0)$. This uses a realization of world-sheet supersymmetry first found in \cite{Waterson:1986ru}. Other realizations were found in \cite{Schellekens:1987ij,Schellekens:1988ag}. In each of these cases one could presumably build a mapping to even self-dual lattices, proving again
that the number of possibilities is finite. But it is not known what the number of ways of realizing world-sheet supersymmetry in terms of free bosons is. If one could prove that this number is finite, we would know that the entire class of covariant lattice theories with chiral spectra is finite. My guess is that it is.


\section{Moduli}\label{Moduli}

From the current perspective the status of moduli in these early works is rather strange. Calabi-Yau compactifactions, first proposed in the end of 1984 \cite{Candelas:1985en}, were known to have moduli. The problems associated with moduli had also already been discussed \cite{Dine:1985he}. Yet none of the 
free boson or fermion four-dimensional string papers discussed above makes any mention of moduli at all. The reason for this omission is clarified by the
following quote from  a paper by D. Gepner \cite{Gepner:1987qi}. This paper appeared in April 1987 and contained the description of a class of four-dimensional string models based 
on interacting CFT's, namely minimal $N=2$ superconformal field theories. We will return to these ``Gepner models" in the next section. The conclusion of the preprint version of Gepner's paper contains the phrase: 
``The compactifications described here are inherently free of the moduli problem". The reason given for this is that any change in the radius of the free bosons
``would break the $N=1$ conformal gauge condition". In the published version of the paper was inversed, and a statement was added that these compactifications,
contrary to what was believed earlier, are in fact Calabi-Yau compactifications in a special point in moduli space. Moving away from that point by giving vacuum expectations to the moduli leads to perfectly valid string compactifications, but which do not admit an exact CFT description.

This is quite analogous to what happens with covariant lattice models. Narain lattice model have moduli; the usual ones of the torus. The requirement of world-sheet supersymmetry imposes a rigid structure on the right-moving lattice. We took it for granted that this fixed all the moduli; indeed, we did not even discuss moduli.  It is
certainly true that all Narain moduli are removed from the massless spectrum. A Narain lattice is a special case of a covariant lattice where $\Gamma_{22,14}$
takes the form $\Gamma_{22,6}\times E_8$, where $E_8$ contains the space-time lattice $D_5$. Massless scalars in any covariant lattice theory must be
vectors in $D_5$. To be massless, these vectors must have a conformal weight $\frac12$ component in the rest of the right-moving lattice, and a 
conformal weight 1 component from the left sector. In a Narain compactification there always exist states with those properties. Some of he roots extending $D_5$ to $E_8$
are vectors of  $D_5$ and have total conformal weight 1. They can be combined with vertex operators $\partial X^I$ from the left sector, or any root of the left lattice. 
The latter may not exist, but the former are always present, and give rise to precisely the expected $6 \times  22$ generic moduli. 

It follows that if $D_5$ is not extended by roots that are $D_5$ vectors, then these canonical moduli are not present in the theory, in agreement with the fact
that the rigidity of the right-moving lattice obstructs all Lorentz rotations. But there can be other moduli. If there are vectors on the lattice 
$\Gamma_{22,14}$ that have the form $(x_L,x_R)$, with $x^2_L=x^2_R=2$, such that $x_R$ is a vector of $D_5$, then these vectors give rise to massless scalars, which may well be moduli. These vectors should not be combinations of roots of the left- and right lattice. In that case they imply an extension of the $D_5$ lattice
to $D_n$, $n>5$, and this implies that the theory is at least partly a torus compactification from a higher dimension. Then at least some of the moduli are canonical
Narain moduli.

What does $(x_L,x_R)$ correspond to in the even self-dual lattice $\Gamma_{88}$? We can decompose vectors on that lattice as $(u,s,t)$, where 
$u$ is a $D_3$ vector, $s$ a vector in $(D_7)^9$ and $t$ lives in the remaining 22-dimensional space. Clearly $t=x_L$ and $u$ is a $D_3$ vector. What
$s$ is depends on how the conformal weight $\frac12$ is decomposed in terms of the $(D_1)^7$ right-moving lattice. There are two options: a vector of one 
of the $D_1$ factors, or four spinors of four distinct $D_1$'s. In the former case $s$ is just a vector of one of the $D_7$ factors, and $(u,s,t)$ is a vector of norm 4
on $\Gamma_{88}$. In the latter case $s$ is a combination of four spinors of four distinct $D_7$ factors. These have norm $\frac74$ each,  so that the total
norm of $(u,s,t)$ is $1+7+2=10$. There is no simple principle that governs the presence or absence of such special norm 4 or norm 10 vectors on ESDL's. 
Hence there may exist chiral covariant lattice theories without any moduli (other than the dilaton), but they would be nearly impossible to find by any known methods.

The main reason moduli are discussed here is that they led, in the beginning of this century, to a famous big number in string theory. It was conjectured 
(based on flux compactifications and many other ingredients) that the
long-standing problem of stabilizing them has a huge number of solutions: the famous number $10^{500}$  \cite{Ashok:2003gk,Douglas:2004zg}. 
As already stated,
this number has nothing to do with the large numbers discussed earlier. It is also not correct to say that the large number of string compactifications found in the eighties is merely a subset of the $10^{500}$. The different string compactifications generically have different chiral properties and hence are in different moduli spaces, although undoubtedly there will be cases that lie in the same moduli space. Roughly speaking the total number of string compactifications is more like
the product of both large numbers, or more precisely (but still not precisely enough) a sum over the number of stabilized vacua for each moduli space. 
Based on what I know today I would expect the
first large number to be considerably smaller than the second. If we write these numbers as $10^{k}$, my best guess at the moment is that $k$ is of order 10 to 50
for the number of moduli spaces, whereas it may be as large as 272.000 \cite{Taylor:2015xtz} for the maximal number of vacua per moduli space. However, in the latter case it should
be noted that the existence none of these moduli-stabilized deSitter vacua is generally accepted. See \cite{Polchinski:2015bea} for a recent discussion and references.

\section{Vacuum scanning}\label{VacScan}


The rest of this paper is about dealing with big numbers. The four-dimensional string constructions that emerged in 1986 almost invited 
writing computer programs scanning these huge spaces. The two constructions most amenable for computerized scanning are free fermion constructions \cite{Kawai:1986va,Antoniadis:1986rn}
and RCFT tensor products. In fact, free fermion constructions can be viewed as a special case of RCFT tensor products, with the Ising model as a building block. There are numerous papers on parts of the free-fermionic landscape,
see for example \cite{Assel:2010wj} and references therein, and  \cite{Dienes:1990ij} for pioneering work on free fermion vacuum scanning.
But what I really mean by RCFT constructions
are tensor products of $N=2$ minimal models, also known as ``Gepner models" \cite{Gepner:1987vz}. These are the main topic of this chapter. Other building blocks can be considered, but outside this area the terrain is largely undeveloped. 
Essential pieces of formalism are missing, and the few attempts at venturing here have not anything spectacularly different or vastly larger. The only cases that have been considered are Kazama-Suzuki models \cite{Kazama:1988qp}, explored in \cite{Font:1989qc}
and permutation orbifolds \cite{Fuchs:1991vu,Maio:2011qn}. There are also many articles on vacuum scanning using orbifold methods, both in heterotic strings and in orientifolds.
The same is true for the geometric approach, where Calabi-Yau manifolds defined a large
scannable class of manifolds.
Here especially the work of Kreuzer and Skarke  \cite{Kreuzer:2000xy} stands out. This section is not intended as a review of vacuum scanning, but only to give some personal reflections based on my own experiences. But many other people contributed to this subject, see {\it e.g.} \cite{Blumenhagen:2006ci,Dienes:2015xua,Cvetic:2009yh,Anderson:2011ns} for various perspectives and further references. Many more references can be found in my review article \cite{Schellekens:2013bpa}.

\subsection{The goal of vacuum scanning}

 I use the term ``vacuum scanning" because I assume that the results of these scans give us some relevant information about actual metastable ground states
 of string theory, usually called ``vacua".  
 Some people question  the existence of such ground states, especially people who do not like the overwhelmingly large
 landscape of vacua that string theory suggests. But if string theory is relevant for our universe, it must contain at least one ``ground state", no matter how
 one chooses to define it, that describes the Standard Model. Vacuum scanning searches for anything that has the same general features, but differences in gauge interactions and matter. If there is a metastable Standard Model 
 ground state, there is no reason why the alternatives should not exist.
 One may prefer to wait for a mathematically rigorous proof of the existence of one or more
 metastable deSitter vacua with gauge symmetries and chiral matter, but that is not how we usually make progress in physics. Then one could also argue that we
 should not even have started thinking about string phenomenology without first developing a full non-perturbative description of string theory.
 
 But why are we doing this?
I want to make it clear that for me at least, and based on our present knowledge, finding the Standard Model 
is {\it not} the goal of vacuum scanning.
This is not an achievable goal at the moment because too many features of the desirable vacuum cannot be taken into account. This includes
moduli stabilization, supersymmetry breaking and the gauge hierarchy. An even more serious problem is vacuum energy, which in any string theory with broken
supersymmetry comes out in Planck units. One may hope that these issues can be factored out of the problem and dealt with separately, but that is just wishful thinking at the moment.  

The
features most likely to survive are gauge symmetries and chiral matter. But even if we get those  right, the second problem is that we do not really have any idea
how much of the landscape we are missing. Any known construction comes with built-in limitations. Since we know that in the end we are dealing with large numbers, the part we are missing in any scan can be huge, and can usually not be estimated. A nice toy example is provided by the discussion in the appendix.
Here a popular construction method, free complex fermions with (anti)-periodic boundary conditions is shown to cover only a fraction of less than $10^{-370}$
of a larger landscape with a size of order $10^{1000}$. This does not prove anything about the actual string landscape, but it serves as a warning.

The other reason for doing vacuum scans is in order to gain statistical information about the landscape. This might enable us one day to postdict certain
features of the Standard Model, and perhaps even make predictions. The fact that these predictions are statistical in nature is not a valid argument against them.
All predictions in physics are based on statistics. There is a chance of about 1 in $10^{10}$ that  the Higgs boson does not exist, despite all observations. Given the
conjectured number of vacua in the landscape, it is not unthinkable that one feature stands out by a statistical factor of much more than $10^{10}$.
Statistical methods were advocated especially by M. Douglas \cite{Douglas:2003um},  However, our present knowledge is still to crude to expect practical 
applications, other than getting a rough idea which features might need new physics, and which ones might just come out right. For some pioneering work
on vacuum scanning and statistics see {\it e.g.} \cite{Gmeiner:2005vz,Dienes:2006ut,Gmeiner:2008xq,Dienes:2007zz}.

At this stage, the real reason for doing vacuum scanning is usually just to get a picture of a previously unexplored region in the landscape. Just as pictures of astrophysical objects, it
is important to get as high a resolution as possible. This allows rare features to become visible, and helps in deciding if all of the features of the Standard Model can
be realized, and  if they can all be realized simultaneously. But just as one should not expect to understand the entire universe from a single high-resolution image, one
should not expect to understand the entire landscape from vacuum scanning of some region. 

\subsection{The computer program Kac}

While working out examples for \cite{Lerche:1986cx} we used a computer program to generate
covariant lattices. The example we produced had chiral spectra, $N=1$ space-time supersymmetry or no supersymmetry, and orthogonal gauge groups.
This computer program has not been preserved. It is therefore hard to say if using modern computers to search more deeply we would be able to
get anything similar to the standard model. However, three family covariant lattice models {\it do} exist, see below.

The foundation for future work on ``vacuum scanning" was laid in the years 1987-1989. I worked with Shimon Yankielowicz \cite{Schellekens:1989am} on properties of fusion rules.  This led to an idea which we called ``simple currents", discovered independently in \cite{Intriligator:1989zw}, and which I will say more about below.
I made a computer program that could work out the Kac-Peterson \cite{KacBook} formula for the modular transformations of affine Lie algebras, and that could compute
their fusion rules using the Verlinde formula \cite{Verlinde:1988sn}.  I made the good decision to preserve these algorithms by making this program future-proof. It has a command line interface and an automatically generated help system, so that it was possible to remember how to run it even long after publication of the paper it was used for. Previous programs I had written were ad-hoc, written for a specific computation only, and would greet the user with messages like ``Enter N", without explaining what ``N" was. 

The program
slowly evolved to an extensive C-program I named ``{\tt Kac}" (an acronym for ``Komputations with algebras and currents"). It can compute spectra of tensor products and
coset models, work out the consequences of field identification \cite{Gepner:1989jq}
and the resolution of their fixed points \cite{Schellekens:1989uf}\cite{Fuchs:1995tq}, compute simple current modular invariants \cite{GatoRivera:1991ru,GatoRivera:1990za,Kreuzer:1993tf}, and the modular transformations of simple current extended tensor products \cite{Fuchs:1996dd}. At a later stage, it was extended to boundary conformal field theory, building on work by the Tor Vergata group \cite{Pradisi:1996yd}, which after several steps culminated in a general formula \cite{Fuchs:2000cm}.  The program incorporates a long list of algorithms developed
in the papers mentioned here and others. 

\subsection{Heterotic Gepner Models}

However the first vacuum scanning project did not use the program {\tt Kac} yet. It was a scan of all simple current modular invariants of Gepner models \cite{Schellekens:1989wx}. 
The spectra this produced had chiral $SO(10)$ or $E_6$ gauge group with $N$ families. To our disappointment we did not find any cases with $N=3$, other than the ``exceptional" example that Gepner had already found \cite{Gepner:1987hi}.  At the time of this work the full set of simple current invariants was not known yet.
Their full classification only became available in 1993 \cite{Kreuzer:1993tf}. Instead, we multiplied single current modular invariants. This was done by multiplying a
set of matrices, each generated by a single current $J$. In the tensor product such a current $J$ is a vector $(J_1,\ldots,J_M)$, where $J_i$ is a simple current of a factor in the
$N=2$ tensor product, and $M$ is the number of factors, determined by the requirement that the total central charge be equal to 9. For an $N=2$ minimal model 
factor with level $k$, there are $4k+2$ choices for $J_i$. This leads to a huge number of choices for  $(J_1,\ldots,J_M)$. On top of this one has to make
such a choice for each factor in the modular matrix product. We considered products of up to 9 matrices. Since this space was too large to explore systematically, we used a stochastic procedure. We made random choices  for the current components $J_i$ in each matrix. We observed that the list of spectra saturated after a while, and assumed that this meant that we had essentially explored the entire set. 

The main outcome of this work was that the number of families in each minimal model combination was quantized, usually in units of 6 and 4. This made
the conclusion that there were no three-family spectra far more convincing. 

The full results of the scan were archived at CERN for the longest time the system allowed, which turned out be 1999. But already many years before that year they
had already disappeared. They are  still available in paper form,
but unfortunately not in scanned or electronic form. This list still turned out to useful later, when we (with Beatriz Gato-Rivera) returned to the problem  \cite{GatoRivera:2010gv} to implement some extra feature: we allowed breaking of world-sheet symmetry in the sector of the theory that maps to a bosonic string. We also allowed  
 the $E_6$ gauge group to be broken to an $SU(3)\times SU(2) \times U(1)$  (times other factors) subgroup. All of these possibilities were already mentioned in the 1989 
 paper, but at that time
running this algorithm 
cost about 80 seconds per spectrum, and only a few spectra were collected. We could only run on the CERN IBM, using quota shared by the entire theory division.
Exceeding these quota implied that the entire theory division would not be able to login anymore, and this was enough reason to abandon the project.
However, already enough spectra had been collected to make the disturbing observation that they alway contained particles with fractional electric  charge. This
observation could be turned into a theorem \cite{Schellekens:1989qb}.

When we returned to the heterotic Gepner models more than 20 years later, the computer program we used in 1989 was not available anymore, but fortunately 
I had meanwhile developed the program {\tt Kac}, which was capable of building $N=2$ minimal models as coset CFT's, work out their field identification,
tensor them and compute the full set of simple current invariants. It also takes into account permutation symmetries of identical factors, and the 
charge conjugation symmetries of each factor. For all of the 168 tensor products of minimal models with $c=9$, this program computes all distinct 
MIPFs in just a few seconds, and hence the entire 1989 computation can be reproduced very quickly.
Since we still had the old results in paper form, we could check our new results, which were
obtained in a completely independent way. They agreed, and this also showed that  for the $N=1$ $E_6$ models (the ones with a Calabi-Yau like spectrum) the 1989 scan was essentially complete.

The new features, broken $E_6$ and worldsheet supersymmetry, could not be scanned completely.  To appreciate how much more difficult this is,  consider
a formula derived  in \cite{GatoRivera:1991ru} for the number of distinct modular invariant partition functions for a simple current group $({\mathbf Z}_p)^k$:
\begin{equation}
N_{\rm MIPF} = \prod_{\ell=0}^{k-1} (1+p^{\ell})
\end{equation}
where $p$ is prime. The analogous formula for discrete factors ${\mathbf Z}_{p^n}$ is unknown; if there are different prime factors with a single power
one gets a factor for each prime. A well-known combination of $N=2$ factors is $(3,3,3,3,3)$, five factors of $N=2$ models with $k=3$. This lies in the moduli space
of the quintic Calabi-Yau manifold. If we leave $SO(10)$ unbroken, this has a simple current group $({\mathbb Z}_{20})^5 \times {\mathbb Z}_{4}$. But if we write this in terms of
characters of the $SU(3)\times SU(2) \times U(1)^2$ subgroup of $SO(10)$ the discrete group enlarges to $({\mathbb Z}_{20})^6 \times {\mathbb Z}_{30}
\times {\mathbb Z}_{3}\times {\mathbb Z}_{2}$. Instead of five ${\mathbb Z_5}$ factors, this has seven. This alone increases the number of possibilities
by a factor $(1+5^5)(1+5^6)=48846876$. This required us to resort to stochastic methods once again.

The results for broken $E_6$ were was again somewhat disappointing. There was an even clearer family number quantization, which cases that previously
only had quanta of 12 or 24 also showing new spectra quantized in units of 6. But there were no three family spectra. To my knowledge the origin of family quantization
in these models has never been clarified.

But then in two subsequent papers we added another ingredient. This was to replace an $N=2$ factor in the bosonic sector of the heterotic string by an entirely different CFT with the same modular transformation properties, a trick we called ``heterotic weight lifting" \cite{GatoRivera:2009yt}. It turns out that this completely removed 
family quantization, and allowed us to get three family spectra in abundance. Family quantization in units of merely 4 and 6 turned out to be an artifact of the left-right 
constructions considered until then, and using heterotic weight lifting and a closely related method applied to the
superfluous $B-L$ symmetry  we explored  \cite{GatoRivera:2010xn}\cite{GatoRivera:2010fi} an entirely novel part of the heterotic landscape. 

As a by-product three family models were found for supercurrents of the form
 $$(v,0,\ldots,0,\sqrt{3},0,\ldots,0)\ .$$ These where obtained as tensor products of $k=1, N=2$ minimal models. These are examples of Gepner models, but with the
 special property that they can be written in terms of free bosons, and hence as covariant lattices. 

\subsection{Gepner Orientifolds}

The roots of my involvement with open string or orientifold model building can also be traced back to   the late eighties:
to  the  paper by Eric Verlinde \cite{Verlinde:1988sn}
leading to the ``Verlinde formula" for fusion rules, the  paper by Cardy \cite{Cardy:1989ir} that related the Verlinde formula to boundary conformal field theory, and my own work
with S. Yankielowicz on simple currents \cite{Schellekens:1989am} as well as our work on fixed point resolution \cite{Schellekens:1989uf}. The link with fixed point resolution was made by the Rome group, as
early as 1991 \cite{Bianchi:1991rd}. During the last decade of last century this link was strengthened, and it became clear that the fixed point resolution matrices
played an essential r\^ole in the computation of boundary reflection coefficients in open string conformal field theory. Meanwhile the formalism for computing
such matrices had been  developed further in \cite{Fuchs:1995zr,Fuchs:1996dd}, and the formalism for computing boundary coefficients for non-diagonal modular 
invariants was worked out in various examples, see {\it e.g.}   \cite{Pradisi:1995qy,Pradisi:1995pp,Petkova:1994zs,Behrend:1999bn}. A condition for the completeness
of boundaries was formulated in \cite{Pradisi:1996yd}.
The fixed point resolution formalism was applied to open string boundaries in \cite{Fuchs:1997kt}. Another ingredient that was needed were crosscap coefficients for orientifold planes, as well as a CFT characterization of all such planes. The foundation for this work was given in another paper by the Rome group 
\cite{Pradisi:1995qy}. We started the generalization of this work to general simple current MIPFs, as well as simple current related Klein bottles in 
\cite{Huiszoon:1999xq,Huiszoon:1999jw}. Meanwhile the boundary state formalism was also generalized to other simple
current MIPFs \cite{Birke:1999ik,Fuchs:1999zi,Fuchs:1999xn}. It was time for a grand synthesis, which was finally achieved in \cite{Fuchs:2000cm}. 

After about a decade of work by many people we had a general formalism that allowed us to compute string  spectra for all simple current MIPFs and for
all simple current related orientifold choices. Furthermore the entire formalism had been built into my computer program {\tt Kac}, in order to check it for consistency,
for example to check the integrality of all partition function coefficients. At the end of the year 2000 this formalism could have been applied straightforwardly to vacuum scanning in Gepner orientifolds. Examples of such orientifolds had already been worked out in 1996 \cite{Angelantonj:1996mw}, but without any aim at Standard Model phenomenology. 
On the other hand, orbifold-orientifold methods starting being used with some success. In \cite{Ibanez:2001nd} the Standard Model was obtained from
intersecting branes, but without requiring stability.  
A remaining challenge was to obtain a stable, supersymmetric realization with full cancellation of all tadpoles. When this still had not been
achieved in 2004 we decided to see if our formalism could make a contribution. 

It would be an exaggeration to say that this was merely a matter of pushing a button, but it is true that all ingredients were already
in place. The main work that was needed were several optimizations in order to make the computations doable. We were facing several large numbers. First of all there are 168 combinations of $N=2$ minimal models. Each has hundreds or thousands of primary fields. We wanted to consider all simple current MIPFs. There are
5403 in total, for the 168 Gepner models combined. For each MIPF there is a handful to a few tens of orientifold choices. This gives a grand total of 49304. 
For each of these one has to compute all possible boundary coefficients. Typically there are hundreds or thousands of boundaries. A standard model
configuration is obtained by combining four sets of boundaries, in such a way that  open strings ending on these boundaries produce a massless
spectrum that corresponds to the Standard Model. 

We aimed for a set of realizations closely related to the one first written down by the Madrid group 
\cite{Ibanez:2001nd}. These ``Madrid models" are build out of four intersecting branes stacks: a $U(3)$ stack  that produces QCD and baryon number, a $U(1)$ stack that yields
lepton number, a third stack to produce the weak interaction gauge group $SU(2)$ and a fourth one providing endpoint for open string that yield the quark and lepton weak singlets. The third stack can be $U(2)$ or $Sp(2)$, and the fourth one $U(1), O(2)$ or $Sp(2)$ (in the latter case there is an additonal $SU(2)_R$ factor in the gauge group).  
In Gepner orientifolds these stacks are realized as RFCT boundary states, but one can describe the resulting configuration using the more intuitive language of intersecting
branes. The boundary states are either unitary, orthogonal and symplectic, and one can count the total number of combinations  that have the required structure, prior to
demanding a particular stack multiplicity. 
In table \ref{OrientifoldTable} we show some counts for the six possible combinations of stacks. 
 The column ``Total number" gives their total, summed over all orientifolds, MIPFs and orientifold choices. The grand total is $4.5\times 10^{19}$. 
 
 \vskip .3truecm
\renewcommand{\arraystretch}{1.1}
\begin{table}[h!]
\begin{center}
\begin{tabular}{|l|l|l|l|l|} \hline
Type	& Total number	& Number searched	& Percentage\\ \hline
USUU (0+6) &	187648179869355108	& 187171389940312068 & 99.75\%	 \\
UUUU (1+7) &	42766246654184825664	& 42730101309436185264 &	99.92\%  \\
USOU (2)	&    35594807811446520	& 21498035622653976	& 60.40\%  \\
UUOU (3)	&    2579563256116048068	& 720412912488220932 &27.93\%	 \\
USSU (4)	&    4486269786712304	& 2792296847030752	& 62.24\% \\
UUSU (5)	&   187648179869355108	& 90192673747778532	&  48.06\%\\ \hline
Total	 & 45761187347637742772	& 43752168618082181524 &	95.61\% \\
\hline
\end{tabular}
\caption{Gepner Orientifold models by type: search statistics \label{OrientifoldTable}}
\end{center}
\end{table}

 All these brane stack combinations must then be subject to two constraints.  If we give them the Standard Model stack multiplicities,
 their total contribution to the dilaton tadpole must be less than or equal to the contribution of the orientifold plane; furthermore their chiral intersections must
 reproduce the Standard Model spectrum. Especially the latter check is computationally expensive. It involves computing annulus coefficients of the form
 \begin{equation}
 \label{AnnCoeff}
 A^i_{ab} = \sum_m \frac{S^i_m R_{am} R_{bm}}{S_{0m}}
 \end{equation}
 Here $a$ and $b$ are the boundary labels, and $m$ is a label of an ``Ishibahi state". These are the closed string states that can propagate between boundary and orientifold
 states given a certain choice of MIPF. These integers $A^i_{ab}$ give the number of times a state with CFT characters $\chi_i$ appears for the given boundaries $a$ and $b$.
 Now one expands the characters and looks for massless states. Summing over these gives the total number of massless states in the open strings stretching between boundary
 states $a$ and $b$. The chirality of these states is also known, and the net chiral states are compared to the Standard Model. The matrix $S$ is the modular transformation
 matrix of the Gepner model under consideration, and the matrices $R$ depend on $S$ and the MIPF. The matrix $S$ can be obtained straightforwardly from the modular 
 transformation matrices of the $N=2$ minimal models making up the Gepner model, but the resulting matrix can be huge. In the worst case, occurring for the tensor
 product $(1,5,82,82)$, it is a $108612 \times 108612$ matrix. Another computational bottle neck is the $P$-matrix \cite{Pradisi:1995qy} that appears in the computation of the Moebius and
 Klein bottle amplitudes. This matrix is defined as  $P=\sqrt{T}ST^2S\sqrt{T}$, where
 $T$ is the generator of the modular transformation $\tau\rightarrow \tau+1$. Since it involves multiplication of large matrices it is preferable to store it, but 
 for a CFT with $106812$ primaries this requires huge amounts of storage space.

Initially, the computation was done  on the cluster of desktops of the Nikhef theory group. Our  goal was to find the first examples of exact, well-defined
(and hence supersymmetric)  open string theories reproducing the 
Standard Model chirally. Although work on other methods (especially orbifold orientifolds) was approaching that goal, it was surprisingly hard to satisfy all conditions,
and all papers before 2004 made some compromises. We found the first example fairly soon after starting the project. It occurred for the tensor product $(3,8,8,8)$. We had
already understood that there was an additional constraint worth imposing, namely the absence of a massless $U(1)$ gauge boson corresponding to $B-L$. The presence
of such an additional $B-L$-photon is in contradiction with experiment, but one could try to appeal to some low-energy Higgs-like mechanism to make it massive. However, there is a known way to give a mass to this gauge boson already at string level, namely axion mixing, which produces a Stueckelberg-like mechanism. We built a check
for axion mixing into the program, and just a few days after the first success we found an example with a massive $B-L$ photon. We worked out a few more examples, and
produced a short paper \cite{Dijkstra:2004ym} about this as soon as possible, because we were feeling the pressure  of the competition. 

It was our intention to really search all cases, or at least push it as far as we could. We succeeded in pushing it very far indeed thanks to the extensive use of
simple current methods. This organizes all of the aforementioned labels into ``orbits". The boundary and Ishibashi states
are labelled by an orbit label $\ell$ and a simple current $J$. The computationally difficult part is the calculation of all relevant matrices $S, R$ and $P$ 
as well as the annulus coefficients between orbits;
the dependence on the current $J$ can be worked out analytically. In the example mentioned above the 108612 labels are organized into just 2646 orbits. This leads to a 
reduction by a factor $50^3$ for the computation of the matrix $P$, and a reduction by a factor 2500 in the required storage space.  We were in the fortunate situation
that Nikhef was building a cluster that was part of the GRID, to be used for LHC data analysis. We were able to use this facility, because LHC had not started yet. We even
made a minor contribution to LHC physics by putting this system to the test. Indeed, we found some bugs. At one point my student Tim Dijkstra, using the WiFi of the hotel he
was staying in during a conference in Madrid,
 had to remove 18000 mail
messages from his mailbox.  These were crash reports sent erroneously by the GRID software. 
But despite the huge computer power, in some cases the simple current reduction simply was not sufficient, and we had to give up. Two of the 168
Gepner combinations were not considered at all (namely (1,5,42,922) and (1,5,43,628)) and for four others we only considered models of types USUU and UUUU. 
The total number and percentage of configurations that was checked is shown in table \ref{OrientifoldTable}.

But this was still not the end of the work. Open string models must satisfy tadpole cancellation constraints, for reasons of
consistency and stability. Before we had even started the search we had decided that we would allow additional branes (a ``hidden sector") to achieve
the tadpole cancellation. These were required not to have a chiral intersection with the Standard Model. So all solutions to the Standard Model spectrum
constraints were checked for the existence of tadpole cancelling hidden sectors. It was clear very quickly that pursuing that until the end was impossible. In 
a given Gepner model there can be thousands of candidate hidden sector branes. Even after imposing the absence of chiral intersections with the Standard Model,
there can be hundreds left. Each such brane can occur with a maximal multiplicity of order 10. So one is confronted with $10^{\rm hundreds}$ of options, that
cannot be searched systematically.  
We were hit directly by the Big Number problem because the computer program experienced some mysterious segmentation faults each time
it was starting to explore a very large hidden sector. We finally discovered that this was due to a message stating something like ``remaining time
.... years; aborting". The number of years, expressed as an unlimited size integer, was so large that it overflowed the text buffer allocated for it.

We developed a variety of strategies aimed at finding at least one solution per Standard Model configuration. We ended up with
about 200.000 distinct solutions. This number was obtained after removing many spectra that were identical to others. Searches of these kind
have large numbers of degeneracies. Some of them can be taken into account in advance to reduce the search effort, some have a known origin, but can be
taken into account most easily {\it a posteriori}, and there are also some unidentified degeneracies that can only be removed by comparing spectra. 
Table \ref{OrientifoldTableTwo} gives the result for each type (types 6 and 7 are the same as 0 and 1, but with massive $B-L$). Column two gives the number
of Standard Model configurations, expressed in column three as a fraction of the total. Column four and five give the same information including tadpole cancellation.
These numbers are prior to removing degeneracies.  

After the paper was published we understood that there was another constraint that should be imposed, namely
the ``K-theory" constraint, related to global anomalies. In \cite{Uranga:2000xp} Angel Uranga presented a convenient formalism using ``probe branes" to take these 
constraints into account. We did this in
 \cite{GatoRivera:2005qd}. It required a rescan of the entire database, and there was a possibility that the number of tadpole solutions would be decimated by the additional 
 constraint. But this did not happen. Quite the contrary, because in a few cases we were able to push the tadpole cancelation search algorithm a bit further we ended with even more solutions,  211634 in total (after removing degeneracies).

\vskip .3truecm
\renewcommand{\arraystretch}{1.1}
\begin{table}[h!]
\begin{center}
\begin{tabular}{|l|l|l|l|l|l|} \hline
Type	& SMs found	& Fraction & Tadpoles solved & Fraction\\ \hline
USUU (0) & 1096682 & $5.9 \times 10^{-12}$ &	215846 & $2.7 \times 10^{-13}$ \\
USUU (6) &  49794 &	$2.7 \times 10^{-13}$ &   4468& $2.4 \times 10^{-14}$\\
UUUU (1) &	 131704  & $3.1 \times 10^{-15}$  &	1280& $3.0 \times 10^{-17}$  \\
UUUU (7) &	  1306 &  $3.1 \times 10^{-17}$ &	 0& 0 \\
USOU (2)	&  9474494  & $1.1 \times 10^{-10}$  &	431633& $4.8 \times 10^{-12}$ \\
UUOU (3)	&  16891580 & $2.3 \times 10^{-11}$	& 12533& $1.7 \times 10^{-14}$\\
USSU (4)	&  16227372 & $5.8 \times 10^{-9}$ & 	978200& $3.5 \times 10^{-10}$\\
UUSU (5)	&  1178970 & $1.3 \times 10^{-11}$ &	5682& $6.3 \times 10^{-14}$ \\ \hline
Total	 &  45051902 & $1.0 \times 10^{-12}$	&1649642 & $3.8 \times 10^{-14}$ \\
\hline
\end{tabular}
\caption{Gepner Orientifold models by type: success rates\label{OrientifoldTableTwo}}
\end{center}
\end{table}

So far these searches were limited to a particular kind of Standard Model realizations, related to the Madrid model. In 
\cite{Anastasopoulos:2006da} we extended the search to essentially any kind of brane configuration with at most four brane stacks.
In  particular we allowed arbitrary choices for the embedding of the Standard Model hypercharge in the brane stack charges. 
We found that the bulk of the possible models belonged to three classes, distinguished by the contribution that the $U(1)$ in the $SU(3)$ stack
makes to the hypercharge. This contribution can be either $\frac16$ (this includes Pati-Salam models and flipped $SU(5)$), or $-\frac13$ 
(including $SU(5)$ GUTs). There is also a class where this fraction is not fixed by the Standard Model particle charges because the 
massless spectrum is entirely built out of unoriented strings. This implies that one can add charge to one open string endpoint and subtract it
at the other end, keeping all particle charges unchanged. 

This scan was considerably harder than the Madrid model scan, because in the latter case we could make use of the fact that the baryon/QCD brane
and the lepton brane have the same intersections with all others. They only differ in their Chan-Paton multiplicities. This made it possible to search
for candidate baryon and lepton branes in one go, essentially reducing the number of nested loops in the Standard Model search from four to three.
For this reason the generalized search could not be pushed as far as the Madrid model we search. We decided to impose a cut-off on the number
of boundary states at 1750. The scan yielded 19345 chirally distinct realizations of the Standard Model. This includes the eight classes already discussed above.
Each of the 19345 classes may contain thousands of explicit models, each with a different non-chiral spectrum. Indeed, the largest class has about 10 million members
(this class has a rather unappealing $U(3)\times Sp(2) \times Sp(6) \times U(1)$ gauge group, with $Sp(6)$ containing a flavor symmetry group). Chirally distinct
models all have a different phenomenology; one could write 16345 separate papers about them.

The results of this scan were stored in order to make them maximally reproducible. It is now possible to display one of the 19345 models by simply giving a unique
spectrum identification number. It is also possible to reload all the models and check for further criteria. This database proved useful in later work on instanton-induced neutrino masses \cite{Ibanez:2007rs}, $SU(5)$-Yukawa couplings and proton decay \cite{Kiritsis:2009sf,Anastasopoulos:2010hu,Anastasopoulos:2011zz} and discrete symmetries \cite{Ibanez:2012wg}.

We also investigated free fermion orientifolds using the same criteria 
\cite{Kiritsis:2008mu}. This search produced no three-family models. We also searched for tachyon-free non-supersymmetric orientifolds 
\cite{GatoRivera:2007yi,GatoRivera:2008zn}. This was only partly succesful.
The combined constraints of getting the Standard Model spectrum, removing all tachyons and cancelling all tadpoles could not be satisfied within the sample we
considered, but there are solutions when any  of the three constraints is dropped. We suspect that with a sufficiently large sample all constraints can be met.

\section{Final thoughts}

Thirty years after the first signs of a huge landscape started emerging, we still have no idea how big it really is. Current estimates range from 0 to $10^{272000}$ \cite{Taylor:2015xtz}.
The lower bound is based on the actual number of deSitter vacua that has been demonstrated to exist in a way that convinces everyone. 

For a recent discussion on the existence of dS vacua, especially in the KKLT construction \cite{Kachru:2003aw},  see \cite{Polchinski:2015bea}. 
This paper comes out in favor of the
conclusions of KKLT, but references to skeptics can also be found here. The fact that I only cite this single paper is because it is a convenient entry point 
to the most recent discussion, and not because I agree with its conclusion; I am simply unable to decide. 
As far as I can tell, few people expect the actual number of dS vacua in string theory to be zero, unless there is an overlooked no-go theorem. Most skeptics
attack specific parts of certain constructions, such as brane-antibrane uplifting in KKLT. But even if they are only partly right, an explicit dS vacuum may be
subject to a very complicated set of constraints that is hard to satisfy explicitly.  For example, suppose the number $10^{272000}$ 
is reduced by a factor $10^{271000}$ due to
constraints that are very unlikely to be satisfied. 
There would still be $10^{1000}$ vacua left, but they would be impossible to find;  it might well be impossible to demonstrate that any exist.

A large landscape is not as problematic as a poorly distributed one.  If it is simply large it might be possible to show by means of statistical 
distributions that the Standard Model has to be in there somewhere. Such a proof would inevitably imply that it is realized a huge number of times. Presumably 
each realization would give different predictions for new physics or additional digits in observed masses and couplings. This would not be most people's dream of the
ultimate theory, but if such an argument is combined  with a proof that gravity requires string theory (or some generalization),  and that string theory implies a
well-distributed landscape of gauge theories, that would be the end of the story. 

But a distribution with huge holes would be the worst possible outcome, even if the total number of vacua is far less than $10^{500}$. We would not be
able to demonstrate the presence of the Standard Model statistically, nor find it explicitly. 

There are also positive sides to big numbers. An obvious one is that a large landscape is required to explain fine-tuning.
A number larger than about $10^{120}$ is needed for the multiverse explanation of the smallness of the 
cosmological constant \cite{Weinberg:1987dv,Bousso:2000xa},  
and a substantial number of vacua is needed to demystify anthropic tunings in the Standard Model parameter space. A less obvious, and rather speculative advantage is
that big numbers may help us make predictions and falsifications.
Big numbers may lead to vastly different landscape probabilities for certain features or regions of the landscape, making
all but a single one of them statistically relevant. A big hierarchy of probabilities may result either from landscape statistics (vacuum counting) but also from multiverse probabilities, see {\it e.g.}  \cite{SchwartzPerlov:2006hi}\cite{Douglas:2012bu}. 

One may even envisage a scenario where anthropic arguments are in peaceful coexistence with essentially unique predictions. Naively one would expect that a landscape
explanation of anthropic fine-tunings always requires a large ensemble of vacua, and hence necessarily a loss of predictive power.  There would not be just one anthropic
vacuum, because that by itself would require a mysterious fine-tuning.
However, if there is a large hierarchy of probabilities the statistically dominant anthropic vacuum may occur extremely  rarely in the multiverse, but it may still be vastly more likely than the next one.
This would lead to a very strong prediction, but it would still  let the landscape play its r\^ole as an explanation for anthropic fine-tunings.   See sections III.F.4 and III.F.5 of
\cite{Schellekens:2013bpa}
for more discussion of this kind of scenario, as well as a serious potential problem, also discussed below.

This may sound like fantasy, but
consider  the results of \cite{Braun:2014lwp} on a class of flux compactifications. These authors find
a huge exponential suppression of the number of vacua when the rank of the gauge group is increased. Let us, inspired by this result, 
suppose that going from rank 4 to rank 5 costs a factor $10^{-1000}$.
If one can find a convincing anthropic argument that {\it at least} rank 4 is required, one would get a very strong prediction of the rank of the gauge group by a combination of
landscape statistics and anthropics. 

Anthropic arguments of this kind have been made in the limited setting of intersecting brane models with at most two stacks 
\cite{Gato-Rivera:2014afa}. Here indeed the Standard Model gauge group clearly stands out among its competitors, on the basis of molecular complexity.  If one adds more stacks,
other options become available, but extra stacks come at a statistical price. However, this cost may not be the extreme exponential fall-off of the previous paragraph.

That is just as well, because here is also a downside to this sort of argument. It may push us deep into the tail of anthropic distributions. Consider
for example the Standard Model gauge group.
One may imagine anthropic arguments for the necessity of QCD and QED, but the weakest link is
the necessity of the weak interactions \cite{Harnik:2006vj}. One could argue that it is needed to provide chirality to protect quark and lepton masses. This might be true is chirality plus a single light scalar boson is
statistically cheaper than having three or more light fermions. This may be against standard lore about technical naturalness, but not against what we know about landscape distributions for fermion masses and Yukawa couplings. It also fits well with the ``just the Higgs" results of LHC (so far). 
However, this assumed statistical advantage of the ``just the Higgs" scenario over alternatives (Dirac masses, composite models, supersymmetry, etc.) will have to
overcome any statistical gain obtained by dropping the weak interactions altogether. If there is a strong exponential rank dependence as suggested above,
the statistical gain due to rank reduction might be something like $10^{1000}$. This may turn apparently unlikely options into the statistically preferred ones.

Another interesting potential consequence of big numbers was recently suggested in  \cite{Taylor:2015xtz}. These authors argue that F-theory with one specific Calabi-Yau fourfold dominates
everything else by factors as large as $10^{3000}$, simply by having the largest number of stabilized vacua (namely about $10^{272000}$).   It predicts that the fundamental
gauge group is $E_8^9 \times F_4^8 \times (G_2 \times SU(2))^{16}$, making a mockery of Grand Unification. 
It also appears to predict large numbers of dark matter sectors with non-abelian interactions, originating from  the same group. If this kind of dark matter has generic observable signatures
this has the potential of falsifying the entire landscape. If this particular F-theory does not contain the Standard Model, but {\it does} contain another anthropic gauge theory, this would also 
falsify the entire landscape. For example, suppose one can get $SU(3)\times SU(2)\times U(1)$, but only with an even number of families. Then the question whether the
exact Standard Model with three families is realized in another part of  the landscape becomes irrelevant. Perhaps for unknown reasons life with two families is
somewhat challenged, but probably not by a factor $10^{-3000}$. So then the landscape would strongly predict  two or four families, in disagreement with the data.

The authors of \cite{Taylor:2015xtz} caution ``we do not claim that we have proved anything here", but if they are right
it would make all attempts at vacuum scanning performed so far utterly irrelevant. Not only would these scans have had no chance of finding ``the" Standard Model
(I took that for granted already), but it would be
highly questionable if anything significant can be learned about the distribution of physical quantities in the landscape by considering statistically insignificant subsets. 

Big numbers could make everything in this paper irrelevant. They could help falsify the string landscape and thereby string theory as well. But they might
also provide the key to its ultimate vindication.

\vskip 2.truecm
\noindent
{\bf Acknowledgements}\vskip.3cm

\noindent
I would like to thank Dieter L\"ust for reading the manuscript. 
This work has been partially 
supported by funding of the Spanish Ministerio de Econom\'\i a y Competitividad, Research Project
FIS2012-38816, and by the Project CONSOLIDER-INGENIO 2010, Programme CPAN
(CSD2007-00042). 
\vskip .2in
\noindent
\noindent
\leftline{ }
\noindent

\appendix

\section{Euclidean ESDL's and free fermions}

In this appendix we show that not all Euclidean even self-dual lattices (ESDL's) can be written in terms of complex free fermions with periodic and anti-periodic boundary conditions. 
The basic idea of the proof is to compute an upper limit to the total number $F_{8k}$ of fermionic theories in $8k$ dimensions, and show that for sufficiently large $k$ this
{\it upper} limit is smaller than the {\it lower} limit on ESDL's.

If one constructs a modular invariant CFT in $8k$ dimensions out of complex free fermions, one obtains a CFT with $8k$ independent free bosons. The set of physical
states of that CFT is described by momentum states of each of these bosons. These form vectors in $8k$ dimensions. Closure of the operator product expansion
requires these vectors to close under addition. Hence they lie on a lattice. Modular invariance requires each of the lattice momentum states to appear with the
same multiplicity as the vacuum state, {\it i.e.} 1. The lattice vectors are equal to $0,\pm\frac12$ or 1 modulo 2, because of the free fermion boundary conditions. 

If follows that the free fermionic ESDL can be built as direct sum of one-dimensional components that are $D_1$ lattices (see \cite{Lerche:1988np}
for a review of lattice constructions). Those lattices have a root lattice
consisting of all the even integers, and four conjugacy classes of weights represented by the one-dimensional vectors $0$, $\pm\frac12$ and $1$. These classes
are usually (by analogy with $D_n$) labelled as $(0), (s), (c)$ and $(v)$ respectively. The $D_1$ root lattice has a basic cell consisting of the interval $[0,2)$, which has volume 2. Its dual lattice, the weight lattice, has volume $\frac12$. 
The full ESDL  is  described by a list of conjugacy classes (called ``glue vectors"  by mathematicians)
of the form
\begin{equation}
(x_1,\ldots,x_{8k})
\end{equation}
where $x_i$ are the conjugacy classes $(0),(v),(s)$ or $(c)$ of $D_1$. These vectors must have integral inner products, even norm, and
they must reduce the size of the unit cell from $2^{8k}$ to 1. 
If these conditions are met we have obtained an ESDL. If we add a glue vector of order $N$ to a lattice, this reduces the volume of the unit cell by a factor $N$. 
For example, to a single $D_1$ root lattice we can add a vector conjugacy class. This has order 2 and reduces the volume of the unit cell to 1. We can also add a spinor conjugacy class. This has order 4 and reduces to volume of the unit cell to $\frac12$. 

If while building up a set of glue vectors  we add one of order 1, then this means that it is dependent on the previous ones and should not be considered. Therefore 
every factor must have at least order 2, and hence reduce the unit cell volume by a factor 2. Therefore there can be at most $8k$ glue vectors.
Then each fermionic ESDL's   in $8k$ dimensions can be specified by an $8k \times 8k$ matrix with entries $(0),(v), (s)$ or $(c)$. Hence an upper limit
on the number of fermionic ESDL's in $8k$ dimensions is
\begin{equation}
\label{GeeKa}
F_{8k} < 4^{64k^2} \equiv G_{8k}
\end{equation}
Although this is an extremely loose upper limit, it is sufficient to show that for sufficiently large $k$ the number $L_{8k}$ defined in (\ref{SiegelTwo}) overwhelms 
$F_{8k}$. To see why, consider the increment of these numbers from $k$ to $k+1$. We will make use of the following exact formula for the Bernoulli numbers
\begin{equation}
\label{StepSize}
\| B_{2n} \| = \frac{2 (2n)!}{(2\pi)^{2n}}\left[ \sum_{\ell=1}^{\infty} (\ell)^{-2n}\right]
\end{equation}
For large $n$ this is extremely well approximated by just the first term in the sum, and furthermore the exact Bernoulli numbers are slightly larger than the approximation. Using this approximation we find
\begin{equation}
\frac{L_{8(k+1)}}{L_{8k}} \gtrapprox \frac{(8k)! (8k+2)! (8k+4)! (8k+6)!}{4^4(2\pi)^{32k+16}}
\end{equation}
On the other hand, the increment of $G_{8k}$ is
\begin{equation}
\frac{G_{8(k+1)}}{G_{8k}} = 4^{64(2k+1)}
\end{equation}
A factorial $p!$ is always larger than $x^p$, for any given $x$ and sufficiently large $p$; the {\it four} factorials in (\ref{StepSize}) are just overkill. Hence the
increment of $L_{8k}$ eventually becomes larger than the increment in the $G_{8k}$, and therefore for sufficiently large $k$ we have $L_{8k} > G_{8k}$. Numerically,
$G_{8k}$ is much larger than $L_{8k}$ for small $k$, but $L_{8k}$ surpasses $G_{8k}$ for $k > 902$.

It possible to arrive at an estimate that is a lot tighter than (\ref{GeeKa}).
For any given lattice we can start by collecting independent basis vectors as follows. First choose a vector with a spinor entry $s$. We can interchange 
$D_1$ factors such that $s$ is the first entry. Now consider all remaining vectors. By making linear combinations we can nullify their first entry. In this set vectors, take one that contains an entry $s$, if there is one. We commute that entry to the second position, and subtract it from the first vector to nullify it on the
second position. 

We can continue doing that until we have a set of $N$ vectors of the form
\begin{eqnarray*}
&(s,0,0,0,\ldots,0,\ldots,x^1_{8k})\\
&(0,s,0,0,\ldots,0,\ldots,x^2_{8k})\\
&(0,0,s,0,\ldots,0,\ldots,x^3_{8k})\\
&\ldots\\
&(0,0,0,0,\ldots,s,\ldots,x^N_{8k}) 
\end{eqnarray*}
Each factor reduces the volume of the lattice by a factor 4. Hence there can be at most $4k$ of them.  After a finite number of steps there are either no
vectors to choose anymore, so that we are finished, or there are none that contain a spinor entry.

In the latter case we continue with classes that consist entirely of vectors. We follow the same procedure.
After another $M$ steps we will run out of new glue vectors.  By permutations and nullifications we bring the $M$ vectors of order 2 into a form where the entries $N+1,\ldots,N+M$ of these $M$ vectors form an $M\times M$
diagonal matrix ${\rm diag}(v,v,v,\ldots,v)$. The remaining entries of the vectors, beyond entry $N+M$ can only be $0$ and $v$. Furthermore we can make linear combinations
of these order-2 vectors with the $N$ spinor classes, so that their entries $N+1,\ldots,N+M$ are either $0$ or $s$. After this procedure we end up with a matrix of the form 
\begin{eqnarray*}
&(s,0,0,0,\ldots,0,0 &|\ y^1_{N+1},\ldots,y^1_{N+M}\ | \ x^1_{N+M+1}, \ldots,x^1_{8k})\\
&(0,s,0,0,\ldots,0,0 &|\ y^2_{N+1},\ldots,y^2_{N+M}\ |\ x^2_{N+M+1}, \ldots,x^2_{8k})\\
&(0,0,s,0,\ldots,0,0 &|\ y^3_{N+1},\ldots,y^3_{N+M}\ |\ x^3_{N+M+1}, \ldots,x^3_{8k})\\
&\ldots\\
&(0,0,0,0,\ldots,0,s &|\ y^N_{N+1},\ldots,y^N_{N+M}\ |\ x^N_{N+M+1}, \ldots,x^N_{8k})\\
&(0,0,0,0,\ldots,0,0 &|\ v,0,0,0,\ldots,0,0\! \ |\ z^{N+1}_{N+M+1},\ldots,z^{N+1}_{8k})\\
&(0,0,0,0,\ldots,0,0 &|\  0,v,0,0,\ldots,0,0\! \ |\ z^{N+2}_{N+M+1},\ldots,z^{N+2}_{8k})\\
&\ldots\\
&(0,0,0,0,\ldots,0,0 &|\  0,0,0,0,\ldots,0,v\! \ |\ z^{N+M}_{N+M+1},\ldots,z^{N+M}_{8k})
\end{eqnarray*}
Here $y$ can take the values $0$ and $s$, $z$ can take the values $0$ and $v$, and $x$ can take all four values.  
The volume of the unit cell equals 1 if $4^N \times 2^M=2^{8k}$, hence 
\begin{equation}\label{TNMEK}
2N+M=8k
\end{equation}
This implies that the upper left block is a square $N\times N$ matrix, and the two other unfixed blocks have $NM$ entries.
Note that for $N=0$ there is no solution, because then $M=8k$ and the $8k$ vectors $v^i$ are all of the form $v^i_j=v\delta_{ij}$.
This is an integer but not even self-dual lattice. It is also easy to see that for $N=1$ there is only one solution, namely the $D_{8k}$ root lattice
with a spinor conjugacy class added (for $k=1$ this is the $E_8$ root lattice).

Without further restrictions, the total number of possibilities for given $N$ and $M$ equals
\begin{equation}
\label{GoodEstimate}
4^{N^2}2^{NM}2^{MN}=4^{N(8k-N)}
\end{equation}
The exponent is maximal for $N=4k$, and then it is equal to $4^{16k^2}$. Since there are $4k$ possible values for $N$, it follows that
the total number of free fermionic theories $F_{8k}$ satisfies
\begin{equation}
F_{8k} < 4k\ 4^{16k^2}
\end{equation}
Numerically, this becomes smaller than $L_{8k}$ for $k \geq 15$.

This count does not include the requirement that the vectors have even norm and integer inner product. We can estimate that
effect as follows. For the first $N$ vectors, about $1/8$ of  randomly chosen vectors will have even norm ({\it i.e.} their last $N+M$ components have norm $\frac74$), and $1/4$ 
of the dotproducts will be integer. 
Hence there is an additional suppression factor
\begin{equation}
\label{RedOne}
8^{-N}4^{-\frac12 N(N-1)}
\end{equation}

This is unfortunately not exact, because it depends on a statistical estimate of distributions of norm and inner products. We can make it a little bit more
precise. 
Numerically, the chance of randomly chosen vectors of $D_1$ conjugacy classes to have norms
$\frac{n}{8} \mod 2$ has a slight dependence on $n$. We find that the values $n=0,4$ occur with a frequency of almost exactly $12.5\%$, but  for 
$n=1, 5$ we find  $13.28\% \approx \frac{17}{128}$, for $n=3, 7$ the fraction is $11.72\%\approx \frac{15}{128}$, for $n=2$ it is $12.55\% \approx \frac{257}{2048}$ and for $n=6$ 
we find $12.45\%\approx \frac{255}{2048}$. It should be possible to derive this analytically.
These values hold if the number of vector entries is sufficiently large, which in practice means 10 or larger. A similar statistical estimate for the inner products of these norm $\frac74$ vectors shows that it is integer in about $24.8\%$ of all cases. Using $.1172$ instead of $\frac18$
and $.248$ instead of $\frac14$ gives and additional suppression factor. For $N=4k$ this suppression ranges from  $10^{-16}$ for $k=11$ to $.002$ for $k=4$. 
We  will not include this in the numerical results. 
This just implies that the actual upper bound is still a few orders of magnitude smaller than what we compute here.

The  $M$ vectors of order 2 have mutually integer inner products, but they have only $50\%$ chance of having even norm. In addition,
their inner product with the first $N$ vectors is integer or half-integer. Hence we get a reduction factor
\begin{equation}
\label{RedTwo}
2^{-M}2^{-NM}
\end{equation}
This has a much smaller impact than (\ref{RedOne}) and hence the error in the suppression is less relevant here.
Putting this all together we get
\begin{equation}
2^{2N(N+M)}2^{-3N}2^{-N(N-1)}2^{-M}2^{-NM}=2^{8k(N-1)-N^2}
\end{equation}
This can be written as
\begin{equation}
\label{LatticeEstimate}
2^{8k(2k-1)-(N-4k)^2}
\end{equation}
This reaches its maximum for $N=4k$, and at the maximum it equals $2^{8k(2k-1)}$. Therefore a highly plausible (though not mathematically rigorous) 
upper bound is 
\begin{equation}
\label{EfKa}
F_{8k} < 4k\ 2^{8k(2k-1)}
\end{equation}
This is smaller than $2L_{8k}$, the lower bound on the number of ESDL's, for $k\geq 8$. For $k=11$, the value of interest of  \cite{Lerche:1986cx}, it is about $10^{558}$. Hence the free fermionic theories yield a fraction of less that $10^{-370}$ of all ESDL's of dimension 88.

Although all free fermionic partition functions that solve  the conditions considered so far are
ESDL's, there will be many degeneracies among them. In particular, we have at our disposal $(8k-N-M)!$ permutations of the last $8k-N-M$ components,
$2^{N}$ inner automorphisms of the last $N$ $D_1$ factors (interchanging $s$ and $c$), $N!$ permutations of the $N$ spinorial glue vectors and
$M!$ permutations of the $M$ vectorial ones. Furthermore there is an unestimable number of distinct ways of choosing different basis vectors for a given lattice 
other then just permutations of basis vectors. On top of that there are other transformations, such as triality rotations within  $D_4$ sublattices (if any exist). But these reductions are all correlated, so it is difficult to make the estimate more precise. Treating the $D_1$ factor permutations, inner automorphisms and spinor class permutations
as if they were uncorrelated gives and additional suppression $\left((4k)!\right)^2 2^{4k}$ for $N=4k$. This is about $10^{122}$ for $k=11$.

We also have some empirical evidence for the amount of overcounting.
For $k=1$ formula (\ref{EfKa}) overcounts the actual number of distinct lattices by a factor 1024; for $k=2$ by a factor $10^{15}$ and for $k=3$ by a factor $10^{36}$.
For $k=4$ the overcount factor is at least $10^{16}$ (because $F_k$ is that much larger than the maximum number of ESDL's). One is inclined to assume that, because of its factorial nature, the 
exponent of the overcount factor increases at least linearly with $k$. This would suggest an overcount reduction by about a factor $10^{200}$ for $k=11$.

We conclude that there are ESDL's which cannot we written in terms of free fermions. In fact, the fraction of ESDL's that {\it can} be written in terms of free
fermions is infinitesimal, and hence the special cases $k=1$ and $k=2$ are  misleading.
For dimension 88, this fraction is smaller than $10^{-370}$, and probably far less than that. For dimension 24, it is not known to me if all Niemeier lattices have a free fermionic description.

\bibliography{Landscape}{}

\begin{thebibliography}{100}

\bibitem{Lerche:1986cx}
W.~Lerche, D.~Lust, and A.~N. Schellekens, ``{Chiral Four-Dimensional Heterotic
  Strings from Selfdual Lattices},'' {\em Nucl.Phys.}, vol.~B287, p.~477, 1987.

\bibitem{Linde:1986fd}
A.~D. Linde, ``{Eternally Existing Selfreproducing Chaotic Inflationary
  Universe},'' {\em Phys.Lett.}, vol.~B175, pp.~395--400, 1986.

\bibitem{Susskind:2003kw}
L.~Susskind, ``{The Anthropic landscape of string theory},'' in {\em Universe
  or Multiverse?} (B.~J. Carr, ed.), pp.~247--266, Cambridge University Press,
  2003.

\bibitem{Schellekens:2008kg}
A.~Schellekens, ``{The Emperor's Last Clothes? Overlooking the String Theory
  Landscape},'' {\em Rept.Prog.Phys.}, vol.~71, p.~072201, 2008.

\bibitem{Schellekens:2013bpa}
A.~Schellekens, ``{Life at the Interface of Particle Physics and String
  Theory},'' {\em Rev.Mod.Phys.}, vol.~85, no.~4, pp.~1491--1540, 2013.

\bibitem{Ashok:2003gk}
S.~Ashok and M.~R. Douglas, ``{Counting flux vacua},'' {\em JHEP}, vol.~0401,
  p.~060, 2004.

\bibitem{Douglas:2004zg}
M.~R. Douglas, ``{Basic results in vacuum statistics},'' {\em Comptes Rendus
  Physique}, vol.~5, pp.~965--977, 2004.

\bibitem{Taylor:2015xtz}
W.~Taylor and Y.-N. Wang, ``{The F-theory geometry with most flux vacua},''
  2015.

\bibitem{Gross:1984dd}
D.~J. Gross, J.~A. Harvey, E.~J. Martinec, and R.~Rohm, ``{The Heterotic
  String},'' {\em Phys.Rev.Lett.}, vol.~54, pp.~502--505, 1985.

\bibitem{Candelas:1985en}
P.~Candelas, G.~T. Horowitz, A.~Strominger, and E.~Witten, ``{Vacuum
  Configurations for Superstrings},'' {\em Nucl.Phys.}, vol.~B258, pp.~46--74,
  1985.

\bibitem{Dixon:1985jw}
L.~J. Dixon, J.~A. Harvey, C.~Vafa, and E.~Witten, ``{Strings on Orbifolds},''
  {\em Nucl.Phys.}, vol.~B261, pp.~678--686, 1985.

\bibitem{Narain:1985jj}
K.~Narain, ``{New Heterotic String Theories in Uncompactified Dimensions},''
  {\em Phys.Lett.}, vol.~B169, p.~41, 1986.

\bibitem{Narain:1986am}
K.~S. Narain, M.~H. Sarmadi, and E.~Witten, ``{A Note on Toroidal
  Compactification of Heterotic String Theory},'' {\em Nucl. Phys.}, vol.~B279,
  p.~369, 1987.

\bibitem{Strominger:1986uh}
A.~Strominger, ``{Superstrings with Torsion},'' {\em Nucl.Phys.}, vol.~B274,
  p.~253, 1986.

\bibitem{Kawai:1986va}
H.~Kawai, D.~C. Lewellen, and S.~H. Tye, ``{Construction of Four-Dimensional
  Fermionic String Models},'' {\em Phys.Rev.Lett.}, vol.~57, p.~1832, 1986.

\bibitem{Lerche:1986he}
W.~Lerche and D.~Lust, ``{Covariant Heterotic Strings and Odd Selfdual
  Lattices},'' {\em Phys. Lett.}, vol.~B187, p.~45, 1987.

\bibitem{Cohn:1986bn}
J.~Cohn, D.~Friedan, Z.-a. Qiu, and S.~H. Shenker, ``{Covariant Quantization of
  Supersymmetric String Theories: The Spinor Field of the Ramond-neveu-schwarz
  Model},'' {\em Nucl. Phys.}, vol.~B278, p.~577, 1986.

\bibitem{Lerche:1986ae}
W.~Lerche, D.~Lust, and A.~N. Schellekens, ``{Ten-Dimensional Heterotic Strings
  From Niemeier Lattices},'' {\em Phys.Lett.}, vol.~B181, p.~71, 1986.

\bibitem{Niemeier}
H.~Niemeier, ``Definite quadratische formen der dimension 24 und diskriminate
  1,'' {\em J. Number Theory}, no.~5, p.~142, 1973.

\bibitem{Dixon:1986iz}
L.~J. Dixon and J.~A. Harvey, ``{String Theories in Ten-Dimensions Without
  Space-Time Supersymmetry},'' {\em Nucl.Phys.}, vol.~B274, pp.~93--105, 1986.

\bibitem{AlvarezGaume:1986jb}
L.~Alvarez-Gaum\'e, P.~H. Ginsparg, G.~W. Moore, and C.~Vafa, ``{An O(16) x
  O(16) Heterotic String},'' {\em Phys.Lett.}, vol.~B171, p.~155, 1986.

\bibitem{Kawai:1986vd}
H.~Kawai, D.~Lewellen, and S.~Tye, ``{Classification of Closed Fermionic String
  Models},'' {\em Phys.Rev.}, vol.~D34, p.~3794, 1986.

\bibitem{Bennett:1986et}
D.~L. Bennett, N.~Brene, L.~Mizrachi, and H.~B. Nielsen, ``{Confusing the
  Heterotic String},'' {\em Phys. Lett.}, vol.~B178, p.~179, 1986.

\bibitem{Schellekens:1992db}
A.~Schellekens, ``{Meromorphic C = 24 conformal field theories},'' {\em
  Commun.Math.Phys.}, vol.~153, pp.~159--186, 1993.

\bibitem{Antoniadis:1985az}
I.~Antoniadis, C.~Bachas, C.~Kounnas, and P.~Windey, ``{Supersymmetry Among
  Free Fermions and Superstrings},'' {\em Phys.Lett.}, vol.~B171, p.~51, 1986.

\bibitem{Antoniadis:1986rn}
I.~Antoniadis, C.~Bachas, and C.~Kounnas, ``{Four-Dimensional Superstrings},''
  {\em Nucl.Phys.}, vol.~B289, p.~87, 1987.

\bibitem{Narain:1986qm}
K.~Narain, M.~Sarmadi, and C.~Vafa, ``{Asymmetric Orbifolds},'' {\em
  Nucl.Phys.}, vol.~B288, p.~551, 1987.

\bibitem{Conway259}
J.~H. Conway and N.~J.~A. Sloane, ``Low-dimensional lattices. iv. the mass
  formula,'' {\em Proceedings of the Royal Society of London A: Mathematical,
  Physical and Engineering Sciences}, vol.~419, no.~1857, pp.~259--286, 1988.

\bibitem{Lerche:1987sk}
W.~Lerche and A.~Schellekens, ``{The Covariant Lattice Construction Of
  Four-Dimensional Strings},'' 1987.

\bibitem{MR1954971}
O.~D. King, ``A mass formula for unimodular lattices with no roots,'' {\em
  Math. Comp.}, vol.~72, no.~242, pp.~839--863 (electronic), 2003.

\bibitem{MR1279061}
M.~Kervaire, ``Unimodular lattices with a complete root system,'' {\em Enseign.
  Math. (2)}, vol.~40, no.~1-2, pp.~59--104, 1994.

\bibitem{Waterson:1986ru}
G.~Waterson, ``{Bosonic Construction Of An N=2 Extended Superconformal Theory
  In Two-Dimensions},'' {\em Phys.Lett.}, vol.~B171, p.~77, 1986.

\bibitem{Schellekens:1987ij}
A.~Schellekens and N.~Warner, ``{Weyl Groups, Supercurrents and Covariant
  Lattices},'' {\em Nucl.Phys.}, vol.~B308, p.~397, 1988.

\bibitem{Schellekens:1988ag}
A.~Schellekens and N.~Warner, ``{Weyl Groups, Supercurrents and Covariant
  Lattices. 2.},'' {\em Nucl.Phys.}, vol.~B313, p.~41, 1989.

\bibitem{Dine:1985he}
M.~Dine and N.~Seiberg, ``{Is the Superstring Weakly Coupled?},'' {\em
  Phys.Lett.}, vol.~B162, p.~299, 1985.

\bibitem{Gepner:1987qi}
D.~Gepner, ``{Space-Time Supersymmetry in Compactified String Theory and
  Superconformal Models},'' {\em Nucl.Phys.}, vol.~B296, p.~757, 1988.

\bibitem{Polchinski:2015bea}
J.~Polchinski, ``{Brane/antibrane dynamics and KKLT stability},'' 2015.

\bibitem{Assel:2010wj}
B.~Assel, K.~Christodoulides, A.~E. Faraggi, C.~Kounnas, and J.~Rizos,
  ``{Classification of Heterotic Pati-Salam Models},'' {\em Nucl.Phys.},
  vol.~B844, pp.~365--396, 2011.

\bibitem{Dienes:1990ij}
K.~R. Dienes, ``{New string partition functions with vanishing cosmological
  constant},'' {\em Phys. Rev. Lett.}, vol.~65, pp.~1979--1982, 1990.

\bibitem{Gepner:1987vz}
D.~Gepner, ``{Exactly Solvable String Compactifications on Manifolds of SU(N)
  Holonomy},'' {\em Phys.Lett.}, vol.~B199, pp.~380--388, 1987.

\bibitem{Kazama:1988qp}
Y.~Kazama and H.~Suzuki, ``{New N=2 Superconformal Field Theories and
  Superstring Compactification},'' {\em Nucl. Phys.}, vol.~B321, p.~232, 1989.

\bibitem{Font:1989qc}
A.~Font, L.~E. Iba\~nez, and F.~Quevedo, ``{String Compactifications and $N=2$
  Superconformal Coset Constructions},'' {\em Phys. Lett.}, vol.~B224, p.~79,
  1989.

\bibitem{Fuchs:1991vu}
J.~Fuchs, A.~Klemm, and M.~G. Schmidt, ``{Orbifolds by cyclic permutations in
  Gepner type superstrings and in the corresponding Calabi-Yau manifolds},''
  {\em Annals Phys.}, vol.~214, pp.~221--257, 1992.

\bibitem{Maio:2011qn}
M.~Maio and A.~Schellekens, ``{Permutation orbifolds of heterotic Gepner
  models},'' {\em Nucl.Phys.}, vol.~B848, pp.~594--628, 2011.

\bibitem{Kreuzer:2000xy}
M.~Kreuzer and H.~Skarke, ``{Complete classification of reflexive polyhedra in
  four-dimensions},'' {\em Adv.Theor.Math.Phys.}, vol.~4, pp.~1209--1230, 2002.

\bibitem{Blumenhagen:2006ci}
R.~Blumenhagen, B.~Kors, D.~Lust, and S.~Stieberger, ``{Four-dimensional String
  Compactifications with D-Branes, Orientifolds and Fluxes},'' {\em
  Phys.Rept.}, vol.~445, pp.~1--193, 2007.

\bibitem{Dienes:2015xua}
K.~R. Dienes, ``{The String Landscape: A Personal Perspective},'' {\em Adv.
  Ser. Direct. High Energy Phys.}, vol.~22, pp.~81--115, 2015.

\bibitem{Cvetic:2009yh}
M.~Cvetic, J.~Halverson, and R.~Richter, ``{Realistic Yukawa structures from
  orientifold compactifications},'' {\em JHEP}, vol.~12, p.~063, 2009.

\bibitem{Anderson:2011ns}
L.~B. Anderson, J.~Gray, A.~Lukas, and E.~Palti, ``{Two Hundred Heterotic
  Standard Models on Smooth Calabi-Yau Threefolds},'' {\em Phys.Rev.},
  vol.~D84, p.~106005, 2011.

\bibitem{Douglas:2003um}
M.~R. Douglas, ``{The Statistics of string / M theory vacua},'' {\em JHEP},
  vol.~0305, p.~046, 2003.

\bibitem{Gmeiner:2005vz}
F.~Gmeiner, R.~Blumenhagen, G.~Honecker, D.~Lust, and T.~Weigand, ``{One in a
  billion: MSSM-like D-brane statistics},'' {\em JHEP}, vol.~0601, p.~004,
  2006.

\bibitem{Dienes:2006ut}
K.~R. Dienes, ``{Statistics on the heterotic landscape: Gauge groups and
  cosmological constants of four-dimensional heterotic strings},'' {\em Phys.
  Rev.}, vol.~D73, p.~106010, 2006.

\bibitem{Gmeiner:2008xq}
F.~Gmeiner and G.~Honecker, ``{Millions of Standard Models on Z-prime(6)?},''
  {\em JHEP}, vol.~0807, p.~052, 2008.

\bibitem{Dienes:2007zz}
K.~R. Dienes and M.~Lennek, ``{Floating correlations on the string
  landscape},'' {\em AIP Conf.Proc.}, vol.~903, pp.~505--508, 2007.

\bibitem{Schellekens:1989am}
A.~Schellekens and S.~Yankielowicz, ``{Extended Chiral Algebras and Modular
  Invariant Partition Functions},'' {\em Nucl.Phys.}, vol.~B327, p.~673, 1989.

\bibitem{Intriligator:1989zw}
K.~A. Intriligator, ``{Bonus Symmetry in Conformal Field Theory},'' {\em Nucl.
  Phys.}, vol.~B332, p.~541, 1990.

\bibitem{KacBook}
V.~Kac, {\em Infinite Dimensional Lie Algebras}.
\newblock Cambridge University Press, 1985.

\bibitem{Verlinde:1988sn}
E.~P. Verlinde, ``{Fusion Rules and Modular Transformations in 2D Conformal
  Field Theory},'' {\em Nucl. Phys.}, vol.~B300, p.~360, 1988.

\bibitem{Gepner:1989jq}
D.~Gepner, ``{Field Identification in Coset Conformal Field Theories},'' {\em
  Phys. Lett.}, vol.~B222, p.~207, 1989.

\bibitem{Schellekens:1989uf}
A.~Schellekens and S.~Yankielowicz, ``{Field Identification Fixed Points In The
  Coset Construction},'' {\em Nucl.Phys.}, vol.~B334, p.~67, 1990.

\bibitem{Fuchs:1995tq}
J.~Fuchs, B.~Schellekens, and C.~Schweigert, ``{The resolution of field
  identification fixed points in diagonal coset theories},'' {\em Nucl.Phys.},
  vol.~B461, pp.~371--406, 1996.

\bibitem{GatoRivera:1991ru}
B.~Gato-Rivera and A.~Schellekens, ``{Complete classification of simple current
  modular invariants for (Z(p))**k},'' {\em Commun.Math.Phys.}, vol.~145,
  pp.~85--122, 1992.

\bibitem{GatoRivera:1990za}
B.~Gato-Rivera and A.~Schellekens, ``{Complete classification of simple current
  automorphisms},'' {\em Nucl.Phys.}, vol.~B353, pp.~519--537, 1991.

\bibitem{Kreuzer:1993tf}
M.~Kreuzer and A.~Schellekens, ``{Simple currents versus orbifolds with
  discrete torsion: A Complete classification},'' {\em Nucl.Phys.}, vol.~B411,
  pp.~97--121, 1994.

\bibitem{Fuchs:1996dd}
J.~Fuchs, A.~Schellekens, and C.~Schweigert, ``{A Matrix S for all simple
  current extensions},'' {\em Nucl.Phys.}, vol.~B473, pp.~323--366, 1996.

\bibitem{Pradisi:1996yd}
G.~Pradisi, A.~Sagnotti, and Y.~Stanev, ``{Completeness conditions for boundary
  operators in 2-D conformal field theory},'' {\em Phys.Lett.}, vol.~B381,
  pp.~97--104, 1996.

\bibitem{Fuchs:2000cm}
J.~Fuchs, L.~Huiszoon, A.~Schellekens, C.~Schweigert, and J.~Walcher,
  ``{Boundaries, crosscaps and simple currents},'' {\em Phys.Lett.}, vol.~B495,
  pp.~427--434, 2000.

\bibitem{Schellekens:1989wx}
A.~N. Schellekens and S.~Yankielowicz, ``{New Modular Invariants For N=2 Tensor
  Products And Four-Dimensional Strings},'' {\em Nucl.Phys.}, vol.~B330,
  p.~103, 1990.

\bibitem{Gepner:1987hi}
D.~Gepner, ``{String Theory On Calabi-Yau Manifolds: The Three Generations
  Case},'' 1987.

\bibitem{GatoRivera:2010gv}
B.~Gato-Rivera and A.~N. Schellekens, ``{Asymmetric Gepner Models:
  Revisited},'' {\em Nucl.Phys.}, vol.~B841, pp.~100--129, 2010.

\bibitem{Schellekens:1989qb}
A.~Schellekens, ``{Electric Charge Quantization in String Theory},'' {\em
  Phys.Lett.}, vol.~B237, p.~363, 1990.

\bibitem{GatoRivera:2009yt}
B.~Gato-Rivera and A.~Schellekens, ``{Heterotic Weight Lifting},'' {\em
  Nucl.Phys.}, vol.~B828, pp.~375--389, 2010.

\bibitem{GatoRivera:2010xn}
B.~Gato-Rivera and A.~Schellekens, ``{Asymmetric Gepner Models II. Heterotic
  Weight Lifting},'' {\em Nucl.Phys.}, vol.~B846, pp.~429--468, 2011.

\bibitem{GatoRivera:2010fi}
B.~Gato-Rivera and A.~Schellekens, ``{Asymmetric Gepner Models III. B-L
  Lifting},'' {\em Nucl.Phys.}, vol.~B847, pp.~532--548, 2011.

\bibitem{Cardy:1989ir}
J.~L. Cardy, ``{Boundary Conditions, Fusion Rules and the Verlinde Formula},''
  {\em Nucl.Phys.}, vol.~B324, p.~581, 1989.

\bibitem{Bianchi:1991rd}
M.~Bianchi, G.~Pradisi, and A.~Sagnotti, ``{Planar duality in the discrete
  series},'' {\em Phys. Lett.}, vol.~B273, pp.~389--398, 1991.

\bibitem{Fuchs:1995zr}
J.~Fuchs, B.~Schellekens, and C.~Schweigert, ``{From Dynkin diagram symmetries
  to fixed point structures},'' {\em Commun.Math.Phys.}, vol.~180, pp.~39--98,
  1996.

\bibitem{Pradisi:1995qy}
G.~Pradisi, A.~Sagnotti, and Y.~S. Stanev, ``{Planar duality in SU(2) WZW
  models},'' {\em Phys. Lett.}, vol.~B354, pp.~279--286, 1995.

\bibitem{Pradisi:1995pp}
G.~Pradisi, A.~Sagnotti, and Y.~S. Stanev, ``{The Open descendants of
  nondiagonal SU(2) WZW models},'' {\em Phys. Lett.}, vol.~B356, pp.~230--238,
  1995.

\bibitem{Petkova:1994zs}
V.~B. Petkova and J.~B. Zuber, ``{On structure constants of sl(2) theories},''
  {\em Nucl. Phys.}, vol.~B438, pp.~347--372, 1995.

\bibitem{Behrend:1999bn}
R.~E. Behrend, P.~A. Pearce, V.~B. Petkova, and J.-B. Zuber, ``{Boundary
  conditions in rational conformal field theories},'' {\em Nucl.Phys.},
  vol.~B570, pp.~525--589, 2000.

\bibitem{Fuchs:1997kt}
J.~Fuchs and C.~Schweigert, ``{A Classifying algebra for boundary
  conditions},'' {\em Phys. Lett.}, vol.~B414, pp.~251--259, 1997.

\bibitem{Huiszoon:1999xq}
L.~Huiszoon, A.~Schellekens, and N.~Sousa, ``{Klein bottles and simple
  currents},'' {\em Phys.Lett.}, vol.~B470, pp.~95--102, 1999.

\bibitem{Huiszoon:1999jw}
L.~Huiszoon, A.~Schellekens, and N.~Sousa, ``{Open descendants of nondiagonal
  invariants},'' {\em Nucl.Phys.}, vol.~B575, pp.~401--415, 2000.

\bibitem{Birke:1999ik}
L.~Birke, J.~Fuchs, and C.~Schweigert, ``{Symmetry breaking boundary conditions
  and WZW orbifolds},'' {\em Adv. Theor. Math. Phys.}, vol.~3, pp.~671--726,
  1999.

\bibitem{Fuchs:1999zi}
J.~Fuchs and C.~Schweigert, ``{Symmetry breaking boundaries. 1. General
  theory},'' {\em Nucl. Phys.}, vol.~B558, pp.~419--483, 1999.

\bibitem{Fuchs:1999xn}
J.~Fuchs and C.~Schweigert, ``{Symmetry breaking boundaries. 2. More
  structures: Examples},'' {\em Nucl. Phys.}, vol.~B568, pp.~543--593, 2000.

\bibitem{Angelantonj:1996mw}
C.~Angelantonj, M.~Bianchi, G.~Pradisi, A.~Sagnotti, and Y.~S. Stanev,
  ``{Comments on Gepner models and type I vacua in string theory},'' {\em
  Phys.Lett.}, vol.~B387, pp.~743--749, 1996.

\bibitem{Ibanez:2001nd}
L.~E. Iba\~nez, F.~Marchesano, and R.~Rabadan, ``{Getting just the standard
  model at intersecting branes},'' {\em JHEP}, vol.~11, p.~002, 2001.

\bibitem{Dijkstra:2004ym}
T.~Dijkstra, L.~Huiszoon, and A.~N. Schellekens, ``{Chiral supersymmetric
  standard model spectra from orientifolds of Gepner models},'' {\em
  Phys.Lett.}, vol.~B609, pp.~408--417, 2005.

\bibitem{Uranga:2000xp}
A.~M. Uranga, ``{D-brane probes, RR tadpole cancellation and K theory
  charge},'' {\em Nucl. Phys.}, vol.~B598, pp.~225--246, 2001.

\bibitem{GatoRivera:2005qd}
B.~Gato-Rivera and A.~Schellekens, ``{Remarks on global anomalies in RCFT
  orientifolds},'' {\em Phys.Lett.}, vol.~B632, pp.~728--732, 2006.

\bibitem{Anastasopoulos:2006da}
P.~Anastasopoulos, T.~Dijkstra, E.~Kiritsis, and A.~N. Schellekens,
  ``{Orientifolds, hypercharge embeddings and the Standard Model},'' {\em
  Nucl.Phys.}, vol.~B759, pp.~83--146, 2006.

\bibitem{Ibanez:2007rs}
L.~Iba\~nez, A.~Schellekens, and A.~Uranga, ``{Instanton Induced Neutrino
  Majorana Masses in CFT Orientifolds with MSSM-like spectra},'' {\em JHEP},
  vol.~0706, p.~011, 2007.

\bibitem{Kiritsis:2009sf}
E.~Kiritsis, M.~Lennek, and A.~N. Schellekens, ``{SU(5) orientifolds, Yukawa
  couplings, Stringy Instantons and Proton Decay},'' {\em Nucl.Phys.},
  vol.~B829, pp.~298--324, 2010.

\bibitem{Anastasopoulos:2010hu}
P.~Anastasopoulos, G.~Leontaris, R.~Richter, and A.~N. Schellekens, ``{SU(5)
  D-brane realizations, Yukawa couplings and proton stability},'' {\em JHEP},
  vol.~1012, p.~011, 2010.

\bibitem{Anastasopoulos:2011zz}
P.~Anastasopoulos, G.~Leontaris, R.~Richter, and A.~N. Schellekens, ``{Avoiding
  disastrous couplings in SU(5) orientifolds},'' {\em Fortsch.Phys.}, vol.~59,
  pp.~1144--1148, 2011.

\bibitem{Ibanez:2012wg}
L.~Iba\~nez, A.~Schellekens, and A.~Uranga, ``{Discrete Gauge Symmetries in
  Discrete MSSM-like Orientifolds},'' {\em Nucl.Phys.}, vol.~B865,
  pp.~509--540, 2012.

\bibitem{Kiritsis:2008mu}
E.~Kiritsis, M.~Lennek, and A.~N. Schellekens, ``{Free Fermion Orientifolds},''
  {\em JHEP}, vol.~0902, p.~030, 2009.

\bibitem{GatoRivera:2007yi}
B.~Gato-Rivera and A.~N. Schellekens, ``{Non-supersymmetric Tachyon-free
  Type-II and Type-I Closed Strings from RCFT},'' {\em Phys. Lett.}, vol.~B656,
  pp.~127--131, 2007.

\bibitem{GatoRivera:2008zn}
B.~Gato-Rivera and A.~Schellekens, ``{Non-supersymmetric Orientifolds of Gepner
  Models},'' {\em Phys.Lett.}, vol.~B671, pp.~105--110, 2009.

\bibitem{Kachru:2003aw}
S.~Kachru, R.~Kallosh, A.~D. Linde, and S.~P. Trivedi, ``{De Sitter vacua in
  string theory},'' {\em Phys.Rev.}, vol.~D68, p.~046005, 2003.

\bibitem{Weinberg:1987dv}
S.~Weinberg, ``{Anthropic Bound on the Cosmological Constant},'' {\em
  Phys.Rev.Lett.}, vol.~59, p.~2607, 1987.

\bibitem{Bousso:2000xa}
R.~Bousso and J.~Polchinski, ``{Quantization of four form fluxes and dynamical
  neutralization of the cosmological constant},'' {\em JHEP}, vol.~0006,
  p.~006, 2000.

\bibitem{SchwartzPerlov:2006hi}
D.~Schwartz-Perlov and A.~Vilenkin, ``{Probabilities in the Bousso-Polchinski
  multiverse},'' {\em JCAP}, vol.~0606, p.~010, 2006.

\bibitem{Douglas:2012bu}
M.~R. Douglas, ``{The string landscape and low energy supersymmetry},'' in {\em
  Strings, Gauge Fields and the Geometry Behind. The Legacy of Max Kreuzer.},
  pp.~261--288, World Scientific, 2012.

\bibitem{Braun:2014lwp}
A.~P. Braun and T.~Watari, ``{Distribution of the Number of Generations in Flux
  Compactifications},'' {\em Phys. Rev.}, vol.~D90, no.~12, p.~121901, 2014.

\bibitem{Gato-Rivera:2014afa}
B.~Gato-Rivera and A.~N. Schellekens, ``{GUTs without guts},'' 2014.

\bibitem{Harnik:2006vj}
R.~Harnik, G.~D. Kribs, and G.~Perez, ``{A Universe without weak
  interactions},'' {\em Phys.Rev.}, vol.~D74, p.~035006, 2006.

\bibitem{Lerche:1988np}
W.~Lerche, A.~Schellekens, and N.~Warner, ``{Lattices and Strings},'' {\em
  Phys.Rept.}, vol.~177, p.~1, 1989.

\end{thebibliography}

\end{document}